\documentclass[acmlarge,screen]{acmart}

\usepackage{booktabs}
\usepackage{multirow}
\usepackage{framed}
\usepackage{colortbl}
\usepackage{xspace}
\usepackage{graphicx}
\usepackage{xcolor}
\usepackage{todonotes}
\usepackage{caption}
\usepackage{subcaption}
\usepackage{tikz}

\newif\ifcomment

\commentfalse

\ifcomment

\newcommand{\gao}[1]{\textcolor[rgb]{0.3,0.8,0.4}{Gao Jie: #1}}
\newcommand{\ken}[1]{\textcolor[rgb]{0.5,0.2,0.2}{Kenny: #1}}
\newcommand{\zsd}[1]{\textcolor[rgb]{0.5,0.1,0.8}{ZSD: #1}}
\newcommand{\simon}[1]{\textcolor[rgb]{0,0.8,0.4}{Simon: #1}}
\newcommand{\cao}[1]{\textcolor[rgb]{0.5,0.8,0.4}{Cao: #1}}

\newcommand{\added}[1]{\textcolor[rgb]{0.2,0.5,0.8}{#1}}
\newcommand{\deleted}[1]{\textcolor[rgb]{0.8,0.8,0.8}{#1}}

\else

\newcommand{\gao}[1]{}
\newcommand{\ken}[1]{}
\newcommand{\zsd}[1]{}
\newcommand{\simon}[1]{}
\newcommand{\cao}[1]{}
\newcommand{\added}[1]{\textcolor{black}{#1}}
\newcommand{\deleted}[1]{}

\fi

\newcommand{\opencoding}{{\emph{Open Coding}}\xspace}

\newcommand{\codeG}{{\emph{Code Granularity}}\xspace}
\newcommand{\textG}{{\emph{Text Granularity}}\xspace}
\newcommand{\para}{{\emph{Paragraph}}\xspace}
\newcommand{\paraS}{{\emph{Paragraph}}\xspace}
\newcommand{\sent}{{\emph{Sentence}}\xspace}
\newcommand{\sentS}{{\emph{Sentence}}\xspace}
\newcommand{\sele}{{\emph{Selective}}\xspace}
\newcommand{\seleS}{{\emph{Selective}}\xspace}
\newcommand{\sh}{{\emph{Short Codes}}\xspace}
\newcommand{\lo}{{\emph{Long Codes}}\xspace}
\newcommand{\mi}{{\emph{Mixed Codes}}\xspace}

\newcommand{\cs}{{\emph{Confidence Score}}\xspace}
\newcommand{\ra}{{\emph{Rank}}\xspace}
\newcommand{\ca}{{\emph{Containing Ability}}\xspace}
\newcommand{\he}{{\emph{Perceived Helpfulness}}\xspace}
\newcommand{\ph}{{\emph{Perceived Helpfulness}}\xspace}

\newcommand{\pt}{{\emph{Perceived Trustworthiness}}\xspace}
\newcommand{\bt}{{\emph{Behavioral Trust}}\xspace}
\newcommand{\cb}{{\emph{Coding Behavior}}\xspace}
\newcommand{\sr}{{\emph{Selecting Rate}}\xspace}
\newcommand{\dt}{{\emph{Decision Time}}\xspace}
\newcommand{\aiqc}{{\emph{AIQCs}}\xspace}
\newcommand{\name}{{\emph{AIcoder}}\xspace}

\newcommand*\circled[1]{\tikz[baseline=(char.base)]{\node[shape=circle,draw,inner sep=0.5pt] (char) {#1};}}

\newcommand{\numberedphrase}[2]{\circled{#1} \textbf{#2}}

\AtBeginDocument{%
  \providecommand\BibTeX{{%
    \normalfont B\kern-0.5em{\scshape i\kern-0.25em b}\kern-0.8em\TeX}}}

\setcopyright{acmcopyright}
\copyrightyear{2023}
\acmYear{2023}
\acmDOI{XXXXXXX.XXXXXXX}

\acmConference[Conference acronym 'XX]{Make sure to enter the correct
  conference title from your rights confirmation emai}{June 03--05,
  2018}{Woodstock, NY}

\acmBooktitle{Woodstock '18: ACM Symposium on Neural Gaze Detection,
 June 03--05, 2018, Woodstock, NY} 
\acmPrice{15.00}
\acmISBN{978-1-4503-XXXX-X/18/06}

\begin{document}


\title[Impact of Human-AI Interaction on User Trust \\ and Reliance in AI-Assisted Qualitative Coding]{\added{Impact of Human-AI Interaction on User Trust and Reliance in AI-Assisted Qualitative Coding}}





\author{Jie Gao}
\affiliation{%
  \institution{Singapore University of Technology and Design}
  \country{Singapore}
}

\author{Junming Cao}
\affiliation{
  \institution{Fudan University}
  \city{Shanghai}
  \country{China}}
\email{21110240004@m.fudan.edu.cn}

\author{Shunyi Yeo, Kenny Tsu Wei Choo}
\affiliation{
  \institution{Singapore University of Technology and Design}
  \city{Singapore}
  \country{Singapore}}


\author{Zheng Zhang, Toby Jia-Jun Li}
\affiliation{
   \institution{University of Notre Dame}
   \country{USA}
   }


\author{Shengdong Zhao}
\affiliation{
   \institution{National University of Singapore}
   \country{Singapore}
   }

\author{Simon Tangi Perrault}
\affiliation{
  \institution{Singapore University of Technology and Design}
  \country{Singapore}
   }
\email{perrault.simon@gmail.com}

\renewcommand{\shortauthors}{Gao et al.}

\begin{abstract}
\added{While AI shows promise for enhancing the efficiency of qualitative analysis, the unique human-AI interaction resulting from varied coding strategies makes it challenging to develop a trustworthy AI-assisted qualitative coding system (\aiqc) that supports coding tasks effectively. We bridge this gap by exploring the impact of varying \emph{coding strategies} on \emph{user trust and reliance on AI}.
We conducted a mixed-methods split-plot $3\times3$ study, involving 30 participants, and a follow-up study with 6 participants, exploring varying \emph{text selection} and \emph{code length} in the use of our \aiqc system for qualitative analysis. Our results indicate that qualitative open coding should be conceptualized as a series of distinct subtasks, each with differing levels of complexity, and therefore, should be given tailored design considerations. We further observed a discrepancy between perceived and behavioral measures, and emphasized the potential challenges of under- and over-reliance on \aiqc systems. Additional design implications were also proposed for consideration.}
\end{abstract}




\maketitle

\section{Introduction}


Qualitative coding is a fundamental tool within qualitative research \cite{corbin2014basics, straws2015basics, charmaz2014constructing}. It is the initial procedure that converts raw data into a format for subsequent stages of analysis.
Despite its importance, coding remains a laborious and frequently repetitive process requiring multiple iterations. In response, researchers have sought to alleviate this labor-intensive task by harnessing Artificial Intelligence (AI), leading to the development of the \textit{AI-assisted Qualitative Coding system (\aiqc)} \cite{marathe2018semi, chen2018using, rietz2021cody, hong2022scholastic, gebreegziabher2023patat}. 
Primarily, AI can facilitate the coding process by offering suggestions informed by past coding annotations, thus prompting users to consider alternative perspectives and rephrase their codes, particularly during the early coding stages. 

\added{On the other hand, trust is fundamental in constructing human-centric AI systems \cite{bach2022systematic, vereschak2021evaluate}. Jiang et al. \cite{jiang2021supporting} have underscored the multitude of factors that foster distrust between humans and AI within the context of qualitative analysis. These include skepticism towards the AI's capability to execute qualitative analysis reliably, noticeable behavioral disparities between humans and AI, the absence of explanations of AI suggestions, and so on.} 

\added{However, we argue that varying human-AI interactions in qualitative analysis, which arise from different coding strategies—a factor that has seemingly been overlooked—pose unique challenges for AI to consistently provide reliable suggestions throughout the \opencoding process. 
Depending on the final objective, coding approaches can vary greatly. For instance, some might opt to code entire paragraphs with a concise code, providing a broad classification for large text segments. Conversely, others might focus on phrases, applying lengthier codes for more detailed insights. Despite similar processes, these strategies yield vastly different depths and scopes in coding outcomes.}

\added{In particular, there is a cascading chain of influence—from the interaction between humans and AI to the user trust and reliance: the \numberedphrase{1}{human-AI interaction} can substantially vary based on the \numberedphrase{2}{coding strategies} employed. It influences the \numberedphrase{3}{users' input}—essentially, the \numberedphrase{4}{training data for the AI}—which consequently impacts both the \numberedphrase{5}{model's performance} and, therefore, the \numberedphrase{6}{quality of AI suggestions}. This chain of influences, in turn, shapes \numberedphrase{7}{humans' perceived trust and reliance} in AI systems. Therefore, our objective is to bridge this gap by examining this influence chain within the context of qualitative analysis. } 




\added{Trust is a concept with many definitions~\cite{vereschak2021evaluate}. 
For this work, we specifically focus on users' behavioral trust (or reliance), perceived trustworthiness and helpfulness of \aiqc. We selected these facets because current trust-related research in \aiqc primarily addresses users' perceived trustworthiness~\cite{jiang2021supporting}. Moreover, discrepancies between perceived and behavioral trust are common in AI-assisted tasks~\cite{dzindolet2003role, vereschak2021evaluate, papenmeier2022s, vorm2018assessing}, highlighting the need for a holistic understanding of trust. Additionally, it's pertinent to investigate whether imperfect AI can still offer valuable helpfulness, considering the innate complexity of achieving perfection in subjective tasks such as qualitative analysis \cite{kocielnik2019will, chen2016challenges, chen2018using}.}

We then operationalize different coding strategies by controlling the coding granularity and introduce two factors: \textG and \codeG. We specifically aim to understand how varying coding granularity influences the following aspects:



RQ1. How does coding granularity impact the model performance of \aiqc?

RQ2. How does coding granularity impact users' \dt and \cb when using \aiqc?

\added{RQ3. How does coding granularity impact users' \bt (i.e., reliance)?}

RQ4. How does coding granularity impact users' \emph{Perceived Trustworthiness} and \emph{Helpfulness} of \aiqc?

RQ5. How does coding granularity impact users' \textit{Subjective Preferences} when using \aiqc?

\added{In response to these research questions, we carried out a split-plot study involving 30 participants, supplemented by a follow-up study with 6 participants. During the main study, participants were given the task of coding texts with varying granularities utilizing individual AI models. We also collected supplemental data wherein users performed the same tasks without access to the AI models for comparative analysis.}

\added{Our findings suggested that qualitative coding should not be perceived as a uniform task, but as a collection of subtasks with varying levels of difficulty. Certain subtasks were found to be more challenging (\para, \lo), whereas others were comparatively simpler (\sh, \mi, \sent, \sele).
An intriguing discrepancy between perceived and behavioral measures emerged from our study: participants indicated higher \ph for more difficult tasks as compared to simpler ones, but exhibited lower \bt; on the contrary, for simpler tasks, participants demonstrated higher \bt but lower \ph. Our study additionally highlighted the potential pitfalls of both under-reliance and over-reliance on \aiqc. Under-reliance might hinder users from fully exploiting the benefits of \aiqc, while over-reliance could lead to ostensibly focused yet shallow outcomes. These factors necessitate careful deliberation in the design of trustworthy \aiqc.}



The contribution of this work is two-fold: 
\added{\begin{itemize}
\item The results of a user study that sheds light on how human-AI interaction in qualitative analysis impacts \aiqc's model performance, user trust, reliance, and perceived helpfulness, with the varying difficulty of \opencoding subtasks.
\item A series of design principles focused on key factors to consider when designing \aiqc to ensure appropriate reliance, trustworthiness and helpfulness.
\end{itemize}}

\section{Background}

\label{sec:QA}
Qualitative coding is a key tool to analyze qualitative data.
Charmaz~\cite{charmaz2014constructing} presented two phases in the Grounded Theory: \textit{Initial Coding} (or \opencoding) and \textit{Focused Coding}. 
Serving as the preliminary step in transitioning from raw data's concrete ideas and concepts to formulating analytic interpretations \cite{charmaz2014constructing}, \textit{Open Coding} involves assigning a summarizing label to varied segments of data, with sizes ranging from a single word to a full paragraph~\cite{charmaz2014constructing, saldana2021coding, corbin2014basics, decuir2011developing}. Subsequently, these labels or codes undergo thorough discussion within a team, leading to the development of a codebook. This codebook comprises a variety of labels/codes correlating with the raw data, thereby facilitating further data analysis \cite{decuir2011developing}.

\begin{figure}[!htbp]
    \centering
    \includegraphics[scale=0.5]{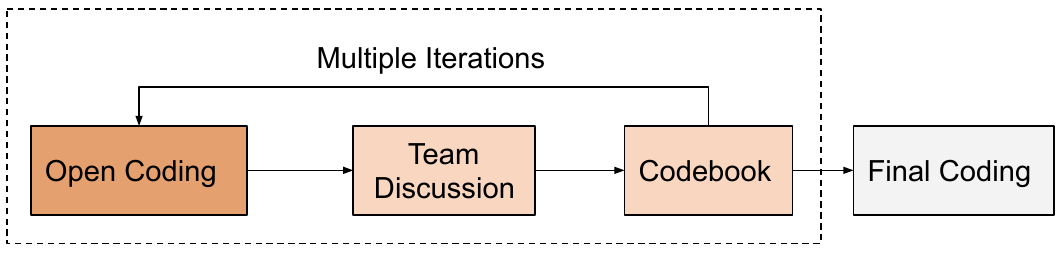}
    \caption{A Circular Coding Process (See More in \cite{decuir2011developing, richards2018practical, saldana2021coding}).}
    \label{fig:codingprocess}
\end{figure}

Nonetheless, the coding process is labor-intensive and demands significant effort, often necessitating multiple iterations within a team \cite{marathe2018semi, rietz2021cody, yan2014semi}. 
This is because the development of the codebook is not a linear process but rather cyclical in nature (see Figure \ref{fig:codingprocess}). 

In particular, \opencoding acts as a key step in this development process \cite{charmaz2014constructing, decuir2011developing} and is inevitably revisited multiple times.
\added{Often, researchers need to revisit raw data, potentially collecting more data and conducting additional coding until reaching saturation, ensuring no nuanced information is overlooked.}

\added{Therefore, the demanding \opencoding process, one of the major causes of slow coding progress, has spurred the development of \aiqc. Our research focuses on this process, exploring how various factors influence AI's ability to provide valuable helpfulness for \opencoding.}


\section{Related Work}

\subsection{\added{AI-assisted Qualitative Coding Systems}}

\added{Many recent studies have explored the application of (semi)automated techniques to facilitate qualitative analysis. Some of these studies \cite{marathe2018semi, crowston2010machine, paredes2017inquire} suggest utilizing code rules for extracting pertinent sections from a given text. For instance, a Boolean rule for \texttt{Definition of arts} may be constructed by linking various keywords using Boolean operators such as AND, OR, and NOT (e.g., \texttt{(definition OR define OR constitute) AND art})~\cite{marathe2018semi}.
This rule is then evaluated against a target text, and a match is identified if their similarity surpasses a specified threshold.}

\added{Moreover, scholars propose code pattern auto-detection in order to support more flexibility \cite{nelson2020computational, gebreegziabher2023patat}. For instance, Nelson's three-step method \cite{nelson2020computational} applies unsupervised machine learning for data pattern discovery, enhancing scalable, exploratory analysis. Meanwhile, PaTAT, introduced by Gebreegziabher et al. \cite{gebreegziabher2023patat}, finds user coding patterns in real-time, predicting future codes. These studies indicate that auto-detection of code rules for partial automation shows significant potential.}


\added{Furthermore, unsupervised machine learning approaches, such as topic modeling~\cite{baumer2017comparing, leeson2019natural, felix2018exploratory, hong2022scholastic}, have proven valuable for detecting topics or labels in qualitative data, especially when dealing with large-scale datasets. By identifying statistical regularities within text, topic modeling can discern thematic patterns, yielding results akin to traditional grounded theory methods \cite{leeson2019natural, muller2016machine}. This approach allows researchers to uncover topics or labels during the early stages of qualitative analysis more effectively~\cite{felix2018exploratory, nguyen2021establishing, kaufmann2020supporting}.}


\added{In addition, supervised techniques such as text classification has gained widespread usage in qualitative analysis. For instance, Yan et al. \cite{yan2014semi} utilized Support Vector Machine (SVM) classification, using pre-selected features and parameters. They trained the SVM model with codes provided by human coders to classify large-scale text data. In a similar vein, Rietz et al.'s Cody \cite{rietz2021cody} used a logistic regression model, employing stochastic gradient descent (SGD) learning. This model was trained to categorize unseen data based on existing annotations.}

\added{While substantial research exists in this field, a comprehensive examination of user reliance, trustworthiness and helpfulness of AI-assisted systems remains scarce \cite{jiang2021supporting}. To our knowledge, this work represents the first attempt to deeply explore how different human-AI interaction strategies within qualitative analysis impact the user trust and reliance for \aiqc.}



\subsection{Trust with \aiqc}
\label{sec:trust}

\added{Trust encompasses various definitions~\cite{vereschak2021evaluate, dzindolet2003role, mayer1995integrative, lee2004trust}. We concentrate on three closely related concepts of trust: \pt, \bt (i.e., reliance), and \ph. 
Firstly, \pt is the perception of a system's trustworthiness, defined as "the extent to which the trustee believes that an automated system will behave as expected"~\cite{papenmeier2022s, vereschak2021evaluate}. We focused on \pt, because according to a study by Jiang et al. \cite{jiang2021supporting}, it was observed that participants generally perceived the trustworthiness of AI in qualitative analysis to be very low.
Secondly, \bt refers to the act of following or accepting someone's recommendation~\cite{dzindolet2003role, vereschak2021evaluate, papenmeier2022s, vorm2018assessing} and is often interchangeable with reliance~\cite{vereschak2021evaluate, dzindolet2003role, lee2004trust}. Our tasks fall under the category of AI-assisted decision making, where AI supports users in reaching their final decisions. As users are the ultimate decision makers, researchers usually measure users' reliance and behavior objectively. Conversely, we chose to study both \bt and \pt because researchers also identified a discrepancy between users' perceived and behavioral trust \cite{vereschak2021evaluate, scharowski2022trust, papenmeier2022s, cao2022understanding}. By observing both subjective and objective aspects, we can gain a more comprehensive understanding of users' overall trust in the system \cite{scharowski2022trust, cao2022understanding, brachman2022reliance}.
Lastly, \ph is described as "the extent to which users perceive the recommendation as being capable of facilitating judgment or decisions" \cite{qin2015perceived, li2013helpfulness}. There is a strong connection between \pt and \he: if users find recommendations helpful, they are more likely to seek advice from those recommendations, thereby fostering trust in the system's capabilities. In addition, we aim to investigate the potential benefits of less trustworthy AI and imperfect system \cite{kocielnik2019will} in subjective tasks \cite{chen2018using, chen2016challenges} and the pitfalls of seemingly perfect and trustworthy systems \cite{banovic2023being}. }

\added{As trust has become a significant topic in the fields of CSCW and HCI \cite{banovic2023being, bach2022systematic, vereschak2021evaluate}, there is also a growing interest in exploring and establishing this element within the \aiqc domain.
In the interview conducted by Jiang et al. \cite{jiang2021supporting}, the authors highlight several sources that contribute to distrust in AI. For instance, they point out discrepancies in the "typical behavior of humans and AI", as AI offers direct suggestions even when humans cannot provide a specific "correct" recommendation, and AI tends to prioritize suggestions with higher probabilities rather than subtle and nuanced insights. They also pointed out that low-precision models often require extra human effort for corrections. Moreover, the absence of explanations from AI and skepticism regarding AI's capacity for creativity and serendipity frequently lead to increased distrust. On the contrary, excessive reliance on AI might prompt researchers to defer to AI as the ultimate authority, potentially compromising human deliberation in the decision-making process. Building on the work of Jiang et al., we delve deeper into trust issues between humans and AI within this domain. We anticipate that in \aiqc, users should establish an appropriate level of trust, avoiding both under- and over-reliance. }

\added{While current \aiqc approaches typically rely on human input for model training and generating code suggestions, there is a lack of research examining the human-AI interactions within the context of \aiqc. These interactions may introduce unique challenges specific to \aiqc that can significantly influence user trust and reliance. Certain interactions have the potential to generate higher-quality input, leading to more accurate code suggestions and improved assistance, while others may have a detrimental effect on these outcomes, subsequently shaping user perceptions and ultimate reliance on the systems.}



\subsection{\added{Human-AI interaction within AIQCs}}
\label{sec:related_work:human_AI_interaction}

\added{There has been significant interest in finding ways for end-users, rather than experts, to interact meaningfully with machine learning systems in order to enhance system performance and user experience \cite{stumpf2009interacting, amershi2014power}. Researchers have also investigated how to refine features of machine learning systems by incorporating human perspectives, particularly in complex qualitative content analysis scenarios \cite{liew2014optimizing}.}
\added{For \aiqc, AI often relies on human-generated training data, which serves as model input. However, unlike many traditional AI tasks, the inherently subjective nature of qualitative analysis poses a unique challenge. This challenge stems from the difficulty of obtaining specific and consistent human inputs, such as labels and text, for AI models. A major contributing factor to this issue is the variability in code granularity present in human coding \cite{saldana2021coding, lindgren2020abstraction}. This variability may lead to inconsistencies and unreliability in the produced data, subsequently affecting the performance and trustworthiness of \aiqc.}

\added{\textG denotes the specific selections of text that are to be coded. Imagine an \opencoding exercise where coding occurs on a word-by-word basis. In such a scenario, the overall context of the text is lacking, making it difficult for the model to suggest any useful codes. Consequently, coders may lose trust in the AI's ability and decide to stop using the system. On the other hand, performing line-by-line or sentence-by-sentence coding would provide the AI with more context. This increased context could potentially enhance the performance of AI models \cite{farra2010sentence}, thereby influencing the system's perceived trustworthiness among users. In this work, we have chosen to examine three different levels of text granularity: \textbf{sentence, paragraph, and selective}. For the last level, users can select phrases of any length, which more closely resembles a regular \opencoding process.}



\added{\codeG refers to the length and specificity of a code \cite{lindgren2020abstraction}. When a code is short, broad, and general (e.g., "experience", "leadership"), the AI might exhibit commendable performance from a classification perspective: the probability of AI suggestions aligning with the user-selected text is elevated, thereby expanding the pool of potential choices within the AI's suggestions. Despite this, dependency on AI assistance under these circumstances is not advisable. There's a risk of users becoming overly dependent on it, which may undermine the depth and diversity of qualitative analysis. On the other side, if users add excessively lengthy or detailed codes (e.g., "he hosts lots of activities", "her pets are very cute"), they may not serve as suitable categories for classification as they could exhibit limited commonality for code reuse. This situation could potentially impact the performance of the AI models. In this work, we have elected to explore three distinct levels of code granularity: \textbf{Short Codes} (i.e., Concise and General Codes),  \textbf{Long Codes} (i.e., Detailed and Comprehensive Codes), and \textbf{Mixed Codes} (Natural Codes). In the case of the latter, users may employ code lengths ranging from one to six words, more closely mirroring a typical \opencoding process.}


\added{Both of the aforementioned scenarios could hinder the AI model from performing as well as expected. As a result, AI could become less useful, leading users to either under-utilize it or rely on it excessively.
Hence, the granularity levels for both codes and text selections must be carefully designed, to enhance the clarity and consistency of the coding scheme~\cite{chen2016challenges, marathe2018semi, rietz2021cody}. We anticipate that our evaluation will distill key insights, thus aiding in the creation of trustworthy, reliable, and beneficial \aiqc.}




\section{AIcoder}

\begin{figure}[!t]
    \centering
    \includegraphics[scale=0.85]{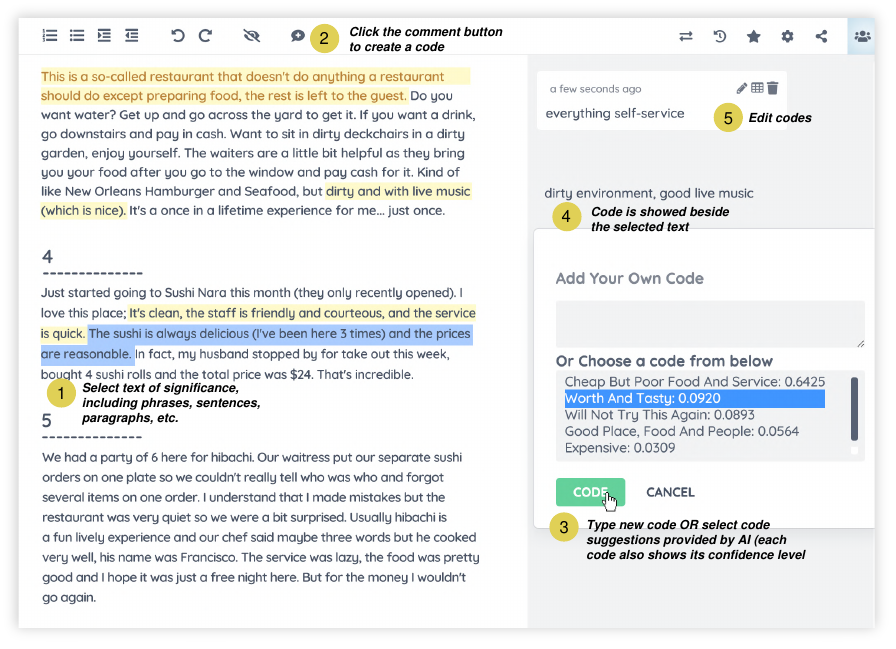}
    \caption{\name Interface. The above figure shows a user was doing coding using \mi. The user can add codes by 1) selecting the text of significance or interest, including phrases, sentences or paragraphs, etc.; 2) clicking the comment button to create a code; 3) typing new code or selecting code suggestions suggested by AI. Each code also shows the confidence level, ranging between 0 and 1; 4) code is shown beside the selected text; 5) edit codes.}
    \label{fig:interface}
\end{figure}

\begin{figure}[htbp]
    \centering
    \includegraphics[scale=2.8]{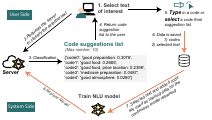}
    \caption{Process of recommendation generation. User side: 1) the user selects the text and clicks on "comment" button; the system 2) automatically requests suggestions from the model server, 3) conducts a classification process, 4) returns a list containing up to 10 code suggestions for the user to either select from or refer to, and 5) the user decides to either create their own codes or select one from the list. System Side: 6) the codes and labeled text are subsequently stored for future use, 7) the selected text and added codes are reused as training data to fine-tune a new model, and 8) the updated model is subsequently deployed onto the server.}
    \label{fig:systembackend}
\end{figure}

We developed a prototype, \name, which adopts an approach similar to prior research that treats qualitative coding as a classification task~\cite{rietz2021cody, yan2014semi}. 
The user interface of \name is displayed in Figure~\ref{fig:interface}, while the backend structure is depicted in Figure~\ref{fig:systembackend}. With \name, users can conveniently highlight any segment of text and assign a specific code. Following the coding, the selected text snippets and their corresponding codes are utilized to fine-tune a model based on their inputs. Ultimately, the user can highlight another piece of text to receive several recommendations from the model. We outline the components of \name in the following sections.

\subsection{Interface} 
The interface (see Figure \ref{fig:interface}) is built on 1) \textit{Etherpad}\footnote{\url{https://etherpad.org/}}, an open source web text editor, supporting users editing text online~\cite{amiryousefi2021impact, bebermeier2019use, goldman2011real}, and 2) its plugin, \textit{ep\_comment\_pages}\footnote{\url{https://github.com/ether/ep_comments_page}}, supporting adding comments beside the text.
With the interface, users can add codes, review code history, modify previous entries, and discard unnecessary codes. In addition, \textit{ep\_comment\_pages} has been customized to provide a list of code suggestions ($n \leq 10$) upon user request. These suggestions are ranked based on their confidence level, which falls within the range of 0 to 1, indicating the cosine similarity score between the predicted labels and the corresponding text \footnote{\url{https://rasa.com/docs/rasa/components/\#dietclassifier}}.

\subsection{AI Model}
\label{sec:AImodel}
The AI model utilized in \name is based on the NLU component of Rasa\footnote{\url{https://rasa.com/docs/rasa/}. Rasa has previously been employed to facilitate conversations in other prototypes within the HCI field \cite{porfirio2019bodystorming}.}, an open-source Python machine learning framework. 
Specifically, we utilize a Rasa-recommended NLU pipeline to train an NLU classification model\footnote{\url{https://rasa.com/docs/rasa/tuning-your-model/\#configuring-tensorflow}}. This includes several components: \texttt{SpacyNLP}, \texttt{SpacyTokenizer}, \texttt{SpacyFeaturizer}, \texttt{RegexFeaturizer}, \texttt{LexicalSyntacticFeaturizer}, two instances of \texttt{CountVectorsFeaturizer}, and \texttt{DIETClassifier}.
Within this pipeline, we selected the pre-trained SpacyNLP language model "en\_core\_web\_trf"\footnote{https://spacy.io/usage/models} to optimize the accuracy of the model.
Additionally, we have chosen the DIET (Dual Intent and Entity Transformer) Classifier \cite{bunk2020diet} to carry out multi-class classification. The processing of the NLU pipeline is performed on a computer running Ubuntu 20.04. This setup includes Tensorflow (2.6.1), CUDA (11.2), and two Nvidia GPU 1080Ti graphics cards. The software stack is completed with Rasa (3.0), Node.js (17.2.0), and MongoDB (5.0.4) installations.


\subsection{Training and Updating}

\subsubsection{Data Saving and Retrieval}
\label{sec:save_data}
The NLU pipeline is trained on each user's individual coding history throughout the coding process. In particular, every user's coded data is independently stored in the database. New proposed codes are then compared with the user's own coding history for being grouped and deduplicated. For instance, if two different sentences receive the same code from one user, they're grouped into a singular "intent" (analogous to the "class" concept in Machine Learning or the "code" in this context) within the Rasa NLU data file: \textit{nlu.yml}. If two similar sentences are coded with two distinct codes by one coder, it is grouped into the "example" for both "intent". Conversely, if two codes carry similar meanings but distinct expressions, they're treated as two separate "intent" for the current version of \name.

\subsubsection{Real-Time Training}
The continuously updated \textit{nlu.yml} file is fed into the NLU pipeline, triggering the automatic training of an updated NLU model. Subsequently, the trained model is immediately uploaded to the Rasa HTTP server, replacing the preceding model.  The entire pipeline typically takes between 10 to 20 seconds, and this duration may increase as the volume of coding data grows. 
To effectively handle user requests and simulate real-time training, we establish two Rasa Open Source servers running on separate ports, utilizing a server-swapping mechanism as a buffering strategy. Specifically, users are able to solicit code suggestions from either of the two servers via HTTP. If \name fails to receive a response from one server due to an ongoing model update process, it promptly switches to the alternate server.

\section{Study Design}
We conducted a user study to assess the impact of various coding strategies on user trust and reliance in \name. To establish different levels of granularity described in section \ref{sec:related_work:human_AI_interaction}, we set parameters for the length of the text selection and the associated code. Users were then asked to undertake qualitative coding at the sentence level, paragraph level, or with more flexible selection within a paragraph. Additionally, they were requested to summarize their codes in either a concise manner (short phrases of no more than three words), a more extended format (phrases containing four to six words), or in a freer style (phrases of mixed lengths ranging from one to six words). \added{These specifications were derived from pilot studies conducted prior to the formal study.} 

Ultimately, we evaluated the model's performance (RQ1 in section \ref{sec:rq:model_performance}); the \dt and \cb (RQ2 in section \ref{sec:rq:coding_behavior}); \added{the \sr and user reliance examination (RQ3 in section \ref{sec:rq:reliance});} participants' self-reported trust in \pt and \he of the system (RQ4 in section \ref{sec:rq:perceived}); as well as their subjective preferences (RQ5 in section \ref{sec:rq:subject}). 

\subsection{Study Task}

\subsubsection{Dataset} We selected the reviews at random from the publicly accessible Yelp reviews dataset\footnote{\url{https://www.yelp.com/dataset/documentation/main}} for our open coding task. We chose Yelp reviews due to two main reasons. First, the content of reviews is a form of text that most people are familiar with, thus facilitating the coding process for participants without imposing significant difficulties. Second, reviews often come in short paragraphs, increasing the likelihood that a code assigned to one paragraph could also apply to another. 

\subsubsection{Pilot Test} In order to ascertain that our participants could complete the open coding tasks within a reasonable timeframe of 1 to 1.5 hours, we conducted a pilot study involving 6 graduate students with proficient English skills. This exercise revealed that a coding task comprising eight paragraphs (with an average of 86.8 words per paragraph) was the most suitable length for our study. 

\begin{figure}[!t]
    \centering
    \includegraphics[scale=1.4]{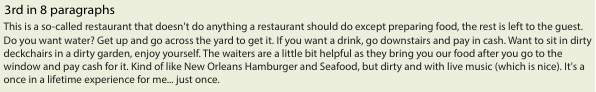}
    \caption{A sample paragraph for the open coding tasks, extracted and preprocessed from the Yelp reviews dataset.}
    \label{fig:task}
\end{figure}

Our pilot study also uncovered several typographical and grammatical errors, along with colloquial references that could potentially hinder participants' understanding. To remedy this, we thoroughly cleaned the text before using it for our open coding task. Figure~\ref{fig:task} depicts one of the revised reviews used in the formal coding tasks.

\subsection{Independent Variables and Conditions}
\begin{figure}[htbp]
    \centering
    \includegraphics[scale=0.61]{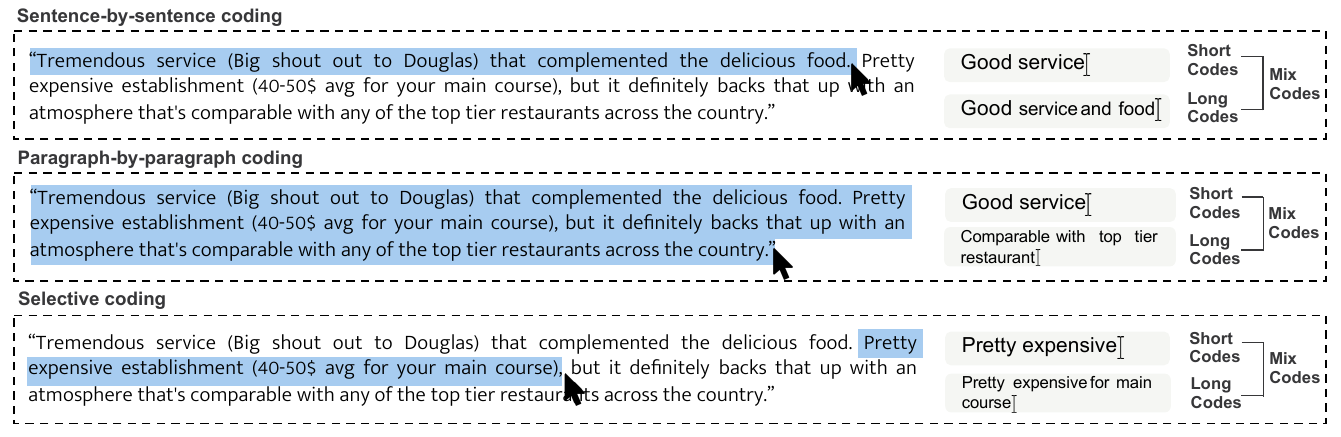}
    \caption{Nine Coding Methods.}
    \label{fig:coding}
\end{figure}

We implemented a split-plot design \cite{lazar2017research} to investigate the effects of two facets of qualitative coding granularity: \emph{Text Granularity} (i.e., unit of analysis or length of text selection) and \emph{Code Granularity} (i.e., length of code in words). The first variable comprises three levels: \sent, \para, and \sele. The second variable also includes three levels: \sh (1-3 words), \lo (4-6 words), and \mi (1-6 words). Consequently, this study encompasses a total of $3\times3 = 9$ conditions (see Table \ref{tab:IVs} and Figure \ref{fig:coding}).

\begin{table}[!htbp]
\caption{Nine conditions corresponding to \textG (i.e., unit of analysis or length of text selection) and \codeG (i.e., length of code in words).}
\label{tab:IVs}
\scalebox{0.8}{\begin{tabular}{@{}cclll@{}}
\toprule
\textbf{} & \multicolumn{1}{l}{} & \multicolumn{3}{c}{\textbf{Text Granularity (text length)}} \\ \midrule
\textbf{} &  & \multicolumn{1}{c}{\begin{tabular}[c]{@{}c@{}}\sent (S)\end{tabular}} & \multicolumn{1}{c}{\begin{tabular}[c]{@{}c@{}}\para (P)\end{tabular}} & \multicolumn{1}{c}{\begin{tabular}[c]{@{}c@{}}\sele (E)\end{tabular}} \\ \midrule
\multirow{3}{*}{\textbf{\begin{tabular}[c]{@{}c@{}}Code\\ Granularity \\ (code length)\end{tabular}}} & \begin{tabular}[c]{@{}c@{}}\sh \\ (1-3 words) (S)\end{tabular} & SS & SP & SE \\
 & \begin{tabular}[c]{@{}c@{}}\lo \\ (4-6 words) (L)\end{tabular} & LS & LP & LE \\
 & \multicolumn{1}{l}{\begin{tabular}[c]{@{}l@{}}\mi \\ (1-6 words) (M)\end{tabular}} & MS & MP & ME \\ \bottomrule
\end{tabular}}
\end{table}

We opted for a mixed-design approach (i.e., a split-plot design) to ensure the experiment duration remained manageable (approximately 1 hour). We designated \emph{Code Granularity} as a between-subject variable to avoid affecting the participants' coding process, particularly their decision-making regarding labels. 

Meanwhile, \textG was counterbalanced according to appearance order across various levels, utilizing a Latin Square method \cite{lazar2017research}.
For each of the three levels of \emph{Text Granularity}, participants were asked to peruse eight selected texts and carry out an open coding task.
The impact of differing text selection lengths can be evaluated by comparing conditions within the same row of Table \ref{tab:IVs}. Conversely, the effect of varying code lengths can be ascertained by comparing conditions within the same column.

\subsection{Participants}

We conducted our study with 30 participants (12 males and 18 females, mean age = 21.9 years old). 
We based our participant selection on the following criteria: 1) age of 18 or above, 2) proficient English reading and writing skills, and 3) enrollment in or completion of an undergraduate programme. 
All participants, being novices in qualitative coding, received appropriate coding training from us prior to the formal study. 
Each participant received compensation for their time equivalent to 7.25 USD per hour, which aligns with the standard rate approved by our institution's IRB.
\added{Additionally, for the follow-up study conducted in section \ref{sec:rq:followup_study}, we recruited 6 participants (3 females, mean age = 26.7 years old) with the same requirements.}

\subsection{Procedure}

We partitioned the 30 participants into three groups of 10, each group assigned to propose either short, long, or mixed codes. Independently of their group, all participants underwent the \sent, \para, and \sele conditions.  
They were instructed to code each sentence, with the option to skip any that were devoid of meaning (\sent); to assign one code to each paragraph (\para); and to code text selections of any length within a paragraph, ranging from phrases to individual or multiple sentences (\sele).

Upon signing the consent form, participants were initially briefed on \opencoding. This was followed by a 15 to 20-minute training session, introducing them to qualitative coding. Subsequently, they began coding tasks under their assigned conditions. \added{They were also given specific research questions to guide their coding process. Primarily, these queries necessitate participants to discern the customers' opinions and attitudes regarding the store or restaurant presented in the coding material.}

\added{To gain insights into users' attitudes towards AI, we implemented a think-aloud protocol during the study. This approach facilitated the observation of participants' coding processes and ensured the tasks were performed correctly. After each study, participants completed a survey and were encouraged to share their reasoning behind the given choices in the survey with the facilitator, and to compare the conditions they experienced. Additionally, we conducted a semi-structured interview at the end of the study to encourage participants to reflect on their experiences.}

\subsection{Dependent Variables}
The study aims to investigate the extent to which users trust, rely on, and find the AI system helpful. 

\subsubsection{Model Performance} \added{To assess the influence of the coding strategies on the model's performance, we employed evaluation metrics from recommendation systems~\cite{tamm2021quality, metrics_accuracy}: in both automatic}\footnote{\added{Automatic Evaluation: We use SentenceTransformers (https://www.sbert.net/) to calculate the similarity between the `ground truth' and the recommendations. A recommendation is deemed relevant if the similarity score meets a predetermined threshold and is considered irrelevant if it falls short of this threshold.}} \added{and human evaluations}\footnote{\added{Human Evaluation: Echoing the automatic evaluation, two authors assigned labels to a subset of code recommendations based on their relevance, designating `0' for `irrelevant' and `1' for `relevant'. They convened twice to establish a shared understanding of the labeling criteria. Upon attaining an inter-rater reliability score (Cohen's kappa $\kappa$) exceeding 0.8, one author proceeded to label the remaining predictions, involving the other author for consultation on more intricate cases.}}, \added{we consider Precision@k and Mean Average Precision (MAP@k), where $k$ signifies the number of suggestions; in automatic evaluations, we also apply Recall@k.}

\added{To facilitate a streamlined evaluation, we limit ourselves to the top five suggestions, denoted as $k=5$. Our observations indicated that the trends for other values of $k$ are similar to those found within the top five.
Furthermore, to facilitate the computation of these metrics, we consider the user's finalized codes as approximately equivalent to the ground truth for each selected text segment\footnote{\added{Computing these metrics in the qualitative coding context requires a ground truth for each text segment. However, qualitative coding is inherently subjective and personal, leading to a lack of consensus on what constitutes a correct suggestion. Two coders might hold conflicting views on one data point. Consequently, we regard each user's final code as the approximate `ground truth' for the respective data. The overlap between each recommendation and this `ground truth' is assessed using both automated techniques and human annotation. However, it's worth noting that this approach could have limitations and introduce measurement errors. We scrutinize these errors in subsequent sections.}}.}

\added{Specifically, $Precision@k (u) = \frac{|rel (u) \cap rec_k(u)|}{k}$ is calculated as the proportion of relevant recommendations among the top k recommendations provided by the system; $Recall@k(u) = \frac{|rel(u) \cap rec_k(u)|}{|rel(u)|}$ is defined as the proportion of relevant recommendations within the top-k recommendations out of the total number of relevant recommendations. 
Likewise, mean $AP@k(u) = \frac{1}{|rec_k(u)|} \sum_{i \in rec_k(u)} \mathbb{I}(i \in rel(u)) Precision@{rank(u,i)}$ is calculated as the average of the Average Precision across all users and requests. This metric incorporates the order information, considering the relevance of items (indicated by $\mathbb{I}(i \in rel(u))$) at their respective ranks (denoted by $rank(u,i)$)}\footnote{\added{$u$ is a user identificator; $i$ is an item identificator; $rec_k(u)$ is a recommendation list for user containing top-k recommended items; $rel(u)$ is a list of relevant items for user $u$ from the test set; $rank(u,i)$ is a position of item $i$ in recommendation list $rec_k(u)$; $I[·]$ is an indicator function.}}.

\subsubsection{Decision Time}
The decision time refers to the duration that users spend on making a decision for each selection, starting from the moment they begin selecting until they finish entering the code. This metric can also serve as an indirect indicator of the difficulty of a coding task \cite{wright1988decision}.

\subsubsection{Coding Behavior}
To enable a thorough comparison across our nine conditions, we examined users' coding behaviors from multiple perspectives, including the number of selections made, the length of selections (in words), the length of codes (in words), and the number of unique codes created. The \cb provides insights into coding strategies employed by participants.

\subsubsection{Behavioral Trust}
Previous studies have utilized user reliance~\cite{vorm2018assessing, papenmeier2022s}, which reflects the willingness to accept system suggestions, as an indicator for \bt. Therefore, we also measure users' reliance~\cite{dzindolet2003role, ashktorab2021ai} by $
\text{\sr} = \frac{\text{Total number of codes selected by users}}{\text{Total number of codes made}}
$. The \sr represents the probability of users selecting a suggested code.

\subsubsection{\emph{\pt}}
We assess users' \pt towards the code suggestions using a five-point Likert scale: 1 = Do not trust at all, 2 = Do not trust, 3 = Neutral, 4 = Relatively trust, 5 = High level of trust.
We adapted our questions from prior research that examined users' trust levels towards classifiers and prediction results~\cite{papenmeier2022accurate, rechkemmer2022confidence, yin2019understanding}, including:

\begin{enumerate}
    \item How much do you trust the \cs of the suggestions?
    \item How much do you trust the \ra of the suggestions?
    \item How much do you trust the system's ability to include your expected code (\ca)?
\end{enumerate}

\added{To promote better understanding of these questions, we verbally illustrate the concepts of the confidence score and ranks to participants as they complete the survey. Furthermore, we provided explicit explanations for each question's intent to participants. For example, we clarified to participants that we were asking whether they believed the confidence score accurately reflected the suggestions' quality; if they viewed the suggestions' ranking as reliable; and if they thought the system was capable of producing their expected codes.}

\subsubsection{\emph{\he}}
Similarly, we include a question on \ph of AI: "How helpful do you think the suggestions were?" Users were also required to provide responses using a five-point Likert scale.

\subsubsection{Subjective Preferences}
We obtained explicit consent from all participants to audio record the entire study. The recorded audio was subsequently transcribed verbatim into text format to facilitate further analysis.

\subsection{Data Analysis}

\subsubsection{Quantitative analysis} We performed statistical analysis~\cite{norman2010likert} using a mixed two-way ANOVA\footnote{\url{https://pingouin-stats.org/generated/pingouin.mixed_anova.html}}: \added{we used repeated measures on \textG, taking into account the random effect of user.} \added{To account for the repeated measures design, we applied appropriate sphericity corrections (Greenhouse-Geisser) when needed, which adjusted both the reported p-values and degrees of freedom when necessary.} Post-hoc comparisons were conducted using pairwise t-tests with Bonferroni correction \added{to account for multiple comparisons}.

\subsubsection{Qualitative analysis} We employed thematic analysis \cite{braun2006using} to derive themes and groupings for qualitative data. Upon becoming acquainted with the data and establishing initial codes, we organized the transcripts into cohesive themes that aligned with the content. 
We reviewed the transcripts and audio recordings to extract pertinent quotes corresponding to each identified theme.

\section{Results}

\begin{table}[!t]
\caption{\added{The model's performance was evaluated through both automatic and human evaluation. Note that the recall@5 metric was only available in the automatic evaluation. All values fall within the range of 0 to 1.}}
\label{tab:model_performance}
\scalebox{0.85}{\begin{tabular}{|c|c|ccc|cc|}
\hline
\multirow{2}{*}{\textbf{\begin{tabular}[c]{@{}c@{}}Factor1: \\ Code Granularity\end{tabular}}} & \multirow{2}{*}{\textbf{\begin{tabular}[c]{@{}c@{}}Factor2: \\ Text Granularity\end{tabular}}} & \multicolumn{3}{c|}{\textbf{Automatic Evaluation}} & \multicolumn{2}{c|}{\textbf{Human Evaluation}} \\ \cline{3-7} 
 &  & \multicolumn{1}{c|}{\begin{tabular}[c]{@{}c@{}}Precision@5\\ (M ± S.D.)\end{tabular}} & \multicolumn{1}{c|}{\begin{tabular}[c]{@{}c@{}}Recall@5\\ (M ± S.D.)\end{tabular}} & \begin{tabular}[c]{@{}c@{}}MAP@5\\ (M ± S.D.)\end{tabular} & \multicolumn{1}{c|}{\begin{tabular}[c]{@{}c@{}}Precision@5\\ (M ± S.D.)\end{tabular}} & \begin{tabular}[c]{@{}c@{}}MAP@5\\ (M ± S.D.)\end{tabular} \\ \hline
\multirow{3}{*}{\begin{tabular}[c]{@{}c@{}}Short Codes \\ (1-3 words)\end{tabular}} & Sentence & \multicolumn{1}{c|}{0.16 ± 0.18} & \multicolumn{1}{c|}{0.30 ± 0.39} & 0.47 ± 0.13 & \multicolumn{1}{c|}{0.20 ± 0.09} & 0.38 ± 0.15 \\ \cline{2-7} 
 & Paragraph & \multicolumn{1}{c|}{0.20 ± 0.13} & \multicolumn{1}{c|}{0.68 ± 0.41} & 0.68 ± 0.23 & \multicolumn{1}{c|}{0.21 ± 0.13} & 0.53 ± 0.31 \\ \cline{2-7} 
 & Selective & \multicolumn{1}{c|}{0.20 ± 0.16} & \multicolumn{1}{c|}{0.47 ± 0.45} & 0.53 ± 0.12 & \multicolumn{1}{c|}{0.21 ± 0.11} & 0.43 ± 0.19 \\ \hline
\multirow{3}{*}{\begin{tabular}[c]{@{}c@{}}Long Codes \\ (4-6 words)\end{tabular}} & Sentence & \multicolumn{1}{c|}{0.16 ± 0.20} & \multicolumn{1}{c|}{0.18 ± 0.19} & 0.42 ± 0.14 & \multicolumn{1}{c|}{0.17 ± 0.08} & 0.37 ± 0.16 \\ \cline{2-7} 
 & Paragraph & \multicolumn{1}{c|}{0.54 ± 0.24} & \multicolumn{1}{c|}{0.65 ± 0.20} & 0.72 ± 0.23 & \multicolumn{1}{c|}{0.44 ± 0.15} & 0.80 ± 0.16 \\ \cline{2-7} 
 & Selective & \multicolumn{1}{c|}{0.42 ± 0.35} & \multicolumn{1}{c|}{0.34 ± 0.29} & 0.57 ± 0.23 & \multicolumn{1}{c|}{0.35 ± 0.24} & 0.56 ± 0.26 \\ \hline
\multirow{3}{*}{\begin{tabular}[c]{@{}c@{}}Mixed Codes \\ (1-6 words)\end{tabular}} & Sentence & \multicolumn{1}{c|}{0.08 ± 0.10} & \multicolumn{1}{c|}{0.20 ± 0.31} & 0.32 ± 0.09 & \multicolumn{1}{c|}{0.12 ± 0.03} & 0.44 ± 0.11 \\ \cline{2-7} 
 & Paragraph & \multicolumn{1}{c|}{0.54 ± 0.30} & \multicolumn{1}{c|}{0.66 ± 0.20} & 0.62 ± 0.24 & \multicolumn{1}{c|}{0.21 ± 0.07} & 0.76 ± 0.20 \\ \cline{2-7} 
 & Selective & \multicolumn{1}{c|}{0.34 ± 0.22} & \multicolumn{1}{c|}{0.44 ± 0.25} & 0.41 ± 0.11 & \multicolumn{1}{c|}{0.15 ± 0.06} & 0.52 ± 0.11 \\ \hline
\end{tabular}}
\end{table}

\subsection{RQ1: Impact on \added{Model Performance}}
\label{sec:rq:model_performance}

\added{The performance of the model for both automatic and human evaluation is reported in Table \ref{tab:model_performance}. Complete details of the statistical analysis can be found in the Appendix \ref{appendix:result:model_performance}.}

\added{The results of both automatic and human evaluation suggest that \codeG does not exert a significant influence on Precision@5, Recall@5, and MAP@5 metrics.}

\added{However, \textbf{\textG does significantly impact Precision@5, Recall@5, and MAP@5 metrics ($p<.01$ for all)}. We noted a trend where \textbf{models performed consistently worse under \sent than \sele and \para} ($p<.05$ for most metrics). Furthermore, \textbf{\para typically outperforms in comparison to \sele}, with $p < .01$ for most differences. Our findings also suggest no interactions between \codeG and \textG.}

\added{For human evaluation, we found no significant effects of \textG on Precision@5. However, \textbf{\textG significantly influences the MAP@5 scores ($p<.001$)}. The trend persists: \textbf{\sent consistently scores lower than both \sele and \para ($p<.01$ in all comparisons)}, and \textbf{\sele consistently lower than \para ($p<.001$)}. Significant interactions between \codeG and \textG were observed under specific conditions. \lo or \mi $\times$ \para significantly outperform \lo or \mi $\times$ \sent, and \mi $\times$ \para also significantly outperform \mi $\times$ \sele.}

\subsubsection{\added{Summary}}

\added{The model, when tasked on \para, demonstrably surpasses both \sele and \sent in both automatic and human evaluations.
The performance boost may be attributed to the verbose text in the \para configuration, which provides a richer data set for the classifier.
However, shorter text selections might inherently convey less information, thereby possibly reducing the probability of a match between users' codes and the selected text. In contrast, longer selections could offer more context, potentially leading to more accurate model classifications. However, this remains speculative, and we seek further validation through users' subjective feedback.}
\subsection{RQ2: Impact on Decision Time and Coding Behavior}
\label{sec:rq:coding_behavior}



\subsubsection{Coding Behavior}
The \cb results are outlined in Table \ref{tab:behavior}. Not all comparisons bear logical consistency—for instance, comparing the length of selections in \para and \sent conditions. 
\added{Further intriguing observations surface when we examine (1) the number and length of selections for \mi, \sh, and \lo under the \sele condition and (2) the difference in code length between \mi vs. \sh and \mi vs. \lo. Specifically:}

\added{(1) Under \sele condition, the length of selections in \lo ($M=29.22$) significantly surpasses that in \mi ($M= 12.07, p<.001$) while no difference between \mi ($M=12.17$) and \sh ($M=12.07$). The number of selections shows no significant difference.}

\added{(2) In the length of code, \lo ($M=5.05$) is longer than \mi ($M=3.37, p<.001$), and \mi ($M=3.37$) is longer than \sh ($M=2.19, p<.001$).} 

\added{(3) In terms of code length, there is no significant difference observed between the codes in \sele ($M=3.13$) and codes \sent ($M=3.19$).}

\begin{table}[!htbp]
\caption{Summary of Coding Behavior. \added{Among nine conditions, \mi $\times$ \sele is the baseline, having no selection and little code constraints and thus closely representing open coding.}}
\label{tab:behavior}
\scalebox{0.8}{\begin{tabular}{|c|c|c|c|c|}
\hline
\multirow{2}{*}{\textbf{\begin{tabular}[c]{@{}c@{}}Factor1: Code\\ Granularity\end{tabular}}} & \multirow{2}{*}{\textbf{\begin{tabular}[c]{@{}c@{}}Factor2: Text\\ Granularity\end{tabular}}} & \multicolumn{3}{c|}{\textbf{Coding Behavior}} \\ \cline{3-5} 
 &  & \textbf{\begin{tabular}[c]{@{}c@{}}Number of \\ Selection\textsuperscript{a}\\ (M±S.D.)\end{tabular}} & \textbf{\begin{tabular}[c]{@{}c@{}}Length of \\ Selection\textsuperscript{b}\\ (M±S.D.)\end{tabular}} & \textbf{\begin{tabular}[c]{@{}c@{}}Length of \\ Code\\ (M±S.D.)\end{tabular}} \\ \hline
\multirow{3}{*}{\begin{tabular}[c]{@{}c@{}}Short Codes\\ (1-3 words)\end{tabular}} & Sentence & -- & -- & 2.06 ± 0.54 \\
 & Paragraph & -- & -- & 2.51 ± 0.59 \\
 & Selective & 29.70 ± 14.09 & 12.17 ± 12.37 & 2.00 ± 0.54 \\ \hline
\multirow{3}{*}{\begin{tabular}[c]{@{}c@{}}Long Codes\\ (4-6 words)\end{tabular}} & Sentence & -- & -- & 4.88 ± 0.92 \\
 & Paragraph & -- & -- & 5.34 ± 0.72 \\
 & Selective & 15.90 ± 7.37 & 29.22 ± 20.71 & 4.93 ± 0.93 \\ \hline
\multirow{3}{*}{\begin{tabular}[c]{@{}c@{}}Mixed Codes\\ (1-6 words)\end{tabular}} & Sentence & -- & -- & 2.63 ± 1.26 \\
 & Paragraph & -- & -- & 5.03 ± 1.08 \\
 & \textbf{Selective (baseline)} & \textbf{28.30 ± 12.51} & \textbf{12.07 ± 9.71} & \textbf{2.46  ± 1.26} \\ \hline
\end{tabular}}

\vspace{1ex}
\raggedright
\footnotesize
\added{\textsuperscript{a} The number of selections for \para is consistently 8, while for \sent is consistently around 35. \\
\textsuperscript{b} The selection length for \para consistently averages around 87 words, while for \sent, it typically averages around 14.6 words.}
\end{table}

\subsubsection{Decision Time}
\begin{figure}[!htbp]
     \centering
     \includegraphics[width=0.5\textwidth]{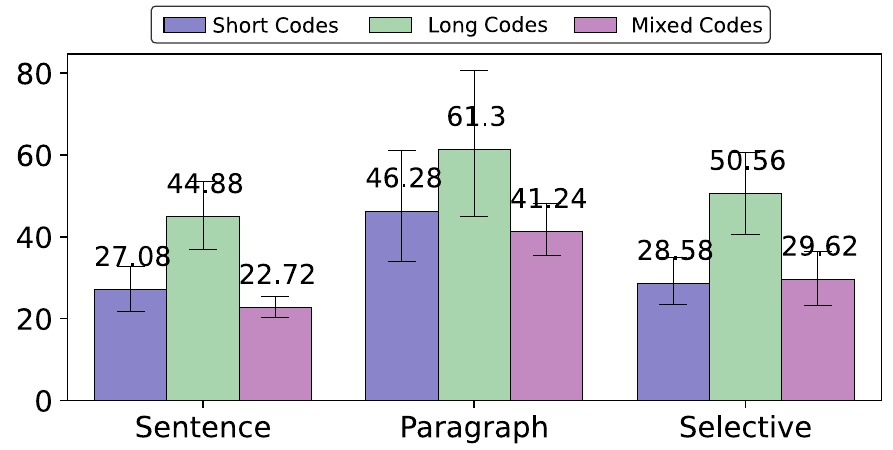}
     \caption{Average Decision Time (Seconds). The time needed to make a decision for each selection. Final results for \sr and \dt. Error bars represent .95 confidence intervals.}
     \label{fig:decision_time}
\end{figure}

Decision Times across all conditions are shown in Figure~\ref{fig:decision_time}.

\paragraph{Code Granularity.}
A significant main effect of \codeG on \dt was detected ($F_{(2,24)} = 11.13, p<.001$). Participants tended to spend more time formulating \lo ($M=52.2s$) compared to \sh ($M=34.0s, p=.014$) and \mi ($M=31.2s, p<.01$).

\paragraph{Text Granularity.}
A significant main effect of \textG on \dt was observed ($F_{(2,48)} = 10.13, p<.001$). Generally, participants required more time to label \paraS ($M=52.2s$) in comparison to \seleS ($M=35.6s, p=.023$) and \sentS ($M=31.6s, p<.001$).

\paragraph{Interactions.} No significant interaction was detected ($p=.97$).

\subsubsection{Summary}
\label{sec:behavior}



\paragraph{Decision Time and Task Difficulty}
Participants found creating \lo more challenging due to a four-word minimum, unlike the one-word minimum for \sh and \mi. Similarly, \dt increased with \textG, with longer coding periods for \para, implying greater task difficulty and time commitment for individual coding selection tasks. 

\paragraph{\emph{\sent $\approx$ \sele? \mi $\approx$ \sh?}}
\label{sec:codingbehavior:sent=sele}
Despite certain disparities, participants demonstrated similar coding behavior across both the \sele and \sent conditions for code length, as well as between \mi and \sh for number and length of selections.


\paragraph{Correlation between Length of Codes and Text}

Crafting longer code names often necessitated larger text selections, a fact further corroborated by the text length between \lo and \sh under \sele condition, while the number of selections significantly decreased in longer codes conditions, while the number of selections in \sh was twice as many.

\subsection{\added{RQ3: Impact on User Reliance}}
\label{sec:rq:reliance}

\added{In this section we examined the users' reliance on AI, we are specifically concerned about the 1) relationship between AI model performance and users' reliance. 2) whether there is a risk of overreliance that could potentially impact coding quality.}

\subsubsection{Selecting Rate}
\begin{figure}[!htbp]
     \centering
     \includegraphics[width=0.5\textwidth]{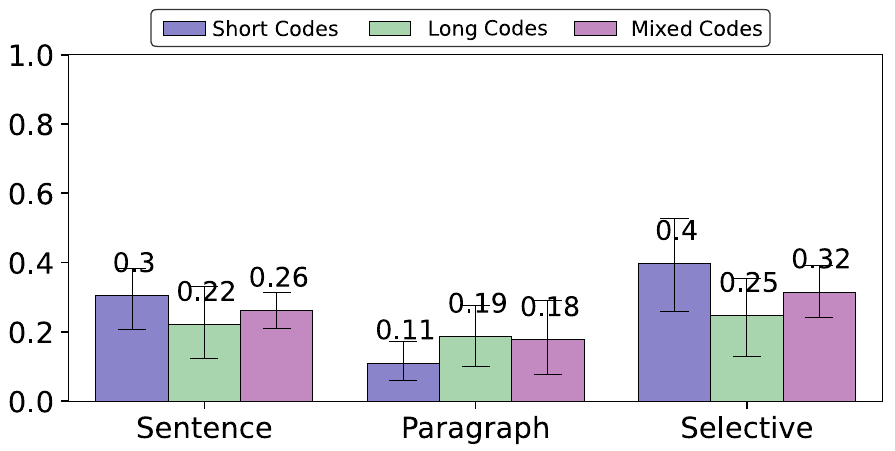}
     \caption{\sr (0-1). Users' receptiveness to code suggestions produced by the system. Final results for \sr and \dt. Error bars represent .95 confidence intervals.}
     \label{fig:selecting_rate}
\end{figure}

The \sr are visualized in Figure \ref{fig:selecting_rate}. Our statistical evaluation reveals that \textbf{\textG has a significant influence on \sr} ($F_{(2, 54)} = 15.838, p<.001$), whereas \codeG does not demonstrate a main effect. 
Pairwise differences are detected (all $p<.05$): \sele registered the highest \sr (with a mean of $M=32\%$ across conditions), followed by \sent (with a mean of $M=26\%$), and lastly \para (with a mean of $M=16\%$).
Moreover, a notable interaction between \emph{Text Granularity} and \emph{Code Granularity} was detected ($F_{(4, 54)} = 2.766, p=.036$). 

The greatest \sr for suggestions was accomplished with \sele $\times$ \sh, in which users chose a suggested code 40\% of the time. This was succeeded by \sele $\times$ \mi with a \sr of 32\%, and \sele $\times$ \lo trailed with a lower rate of 25\%. The \sr pattern remained consistent for \sent coding, with rates fluctuating between 22\% and 30\%. Conversely, \para recorded the least selection rates, which ranged from 11\% to 19\% overall.

Overall, \textbf{user reliance on AI is more pronounced in the \sele condition}, where users are tasked with selecting only pertinent text portions—a situation that closely resembles real-world coding scenarios. 
On the contrary, \textbf{the \para condition had the lowest \sr}, particularly with \sh, likely due to the difficulty of summarizing and coding an entire paragraph with a 3-word limit. This demanding task caused both AI and participants to struggle.

\subsubsection{\added{Correlation between AI Model Performance and Users' Reliance}}
\label{sec:rq:reliance_corr}

\begin{figure}[!t]
     \centering
     \begin{subfigure}[b]{0.4\textwidth}
         \centering
         \includegraphics[width=\textwidth]{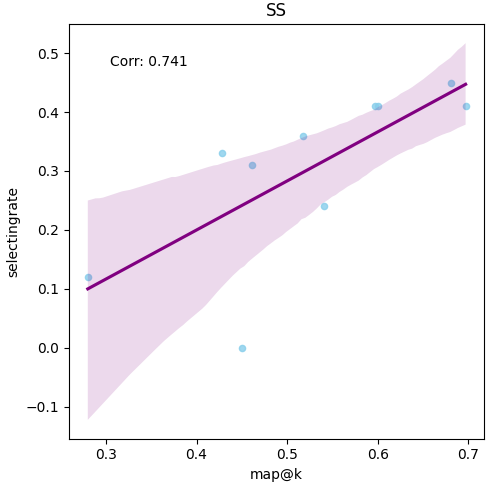}
         \caption{\added{Correlation between selection rate and MAP@5 for \sh $\times$ \sent.}}
         \label{fig:corr:ss}
     \end{subfigure}
     \hfill
     \begin{subfigure}[b]{0.4\textwidth}
         \centering
         \includegraphics[width=\textwidth]{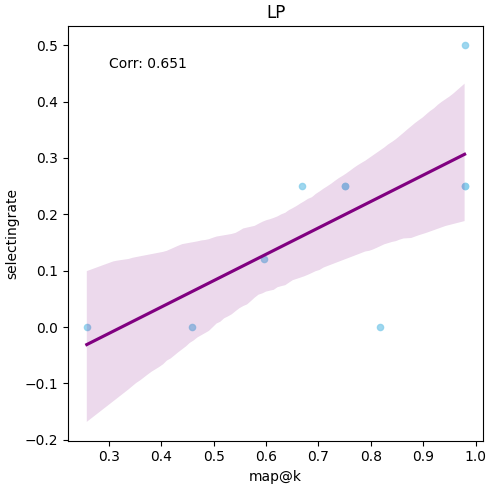}
         \caption{\added{Correlation between selection rate and MAP@5 for \lo $\times$ \para.}}
         \label{fig:corr:lp}
     \end{subfigure}
     \caption{\added{Significant correlations ($p<.05$) for both human and automatic evaluations.}}
\end{figure}

\added{We have identified significant correlations ($p<.05$) for both human and automatic evaluations under two conditions: \sh $\times$ \sent (Figure \ref{fig:corr:ss}), and \lo $\times$ \para (Figure \ref{fig:corr:lp}). In these situations, users displayed an increased \sr as MAP@k increased. This implies that during the coding process of our study, users garnered more assistance from the system in their decision-making as the performance increased.}

\added{We should note, however, that we have used the users' final codes as approximations of `ground truth'. A potential `measurement error' could arise, indicating the possibility of `overreliance'. This is because users might have accepted or made minor modifications after selecting the suggestions--even though they are not the best choice. This could conceivably lead to inflated performance metrics, subsequently resulting in an increase in the MAP@k value (see more details in the acknowledged limitation in section \ref{limitation}).}

\added{Given the strong correlation between \sr and model performance, we are concerned that this effect could be more pronounced, suggesting that some coders may have overly relied on the system under these conditions, thereby affecting the final coding quality. This concern is elevated when observing the specific \sr for each coder: 6 out of 30 participants demonstrated a \sr above 50\%. When reviewing the \sr of these participants under the \sent conditions, a similar trend emerged.}

\subsubsection{\added{Comparing the Coding Results With and Without AI Assistance}}
\label{sec:rq:followup_study}

\added{\paragraph{Supplementary Study} In order to validate our concern, we carried out a supplementary study. This involved 6 more participants performing coding tasks without AI assistance under conditions suspected to foster over-reliance. The procedure is identical to that of our primary study.}

\added{The experimental setup encompasses not only the two previously mentioned conditions that demonstrated a robust correlation between model performance and AI but also a condition that serves as the principal baseline (\mi $\times$ \sele). Additionally, we opted not to include the condition with the highest \sr (\sh $\times$ \sele) due to the analogous behavior observed between users for \sele and \sent, \mi and \sh (see section \ref{sec:behavior}). This similarity led us to anticipate that an analysis of the \sh $\times$ \sent and \mi $\times$ \sele condition would probably yield results comparable to those from the \sh $\times$  \sele condition.}

\added{\paragraph{Results} We decided to perform a qualitative comparison between the final quality of the coding results with and without AI assistance. The results are detailed in Table \ref{tab:comparison_code_results}.}

\added{We observed that codes in \sh $\times$ \sent with AI assistance seem to have lesser code variances than coding without AI, even though their primary category is similar. For instance, `bad food' is a code with AI assistance, while codes such as `cold, old fries' or `dislike burger set' serve as analogs of `bad food' without AI assistance, albeit with more variance. When assisted by AI, users might be presented with a `bad food' suggestion, which they may subsequently adopt instead of proposing other expressions.}

\added{The \sh $\times$ \sent condition appears `acceptable', with participants exhibiting a relatively commendable \sr. Likewise, the \mi $\times$ \sele condition reveals a similar change in code results. We might anticipate similar decreases in other conditions like \sh $\times$ \sele. Interestingly, the \lo $\times$ \para condition seems to demonstrate a relative consistency, regardless of the presence or absence of AI.}

\added{Indeed, the decrease in code variance, to a certain extent, could be perceived as beneficial since it might result in more focused coding and reduce the effort needed to group variances.  \textbf{However, it risks yielding coding outcomes that appear less substantial and somewhat superficial. This could potentially influence the discussion and creation of a codebook in subsequent stages of real qualitative analysis.}}

\begin{table}[!t]
\caption{\added{Comparison of typical coding results for each condition (LP = \lo $\times$ \para, SS = \sh $\times$ \sent, and ME = \mi $\times$ \sele), both with and without the application of AI.  Each cell presents codes derived from a typical user's results under the specified condition. While the "with AI" condition may yield codes with a higher selection rate, they may lack nuanced detail. In contrast, the "without AI" condition tends to generate more detailed codes.}}
\label{tab:comparison_code_results}
\scalebox{0.65}{\added{\begin{tabular}{|l|ll|lll|}
\hline
\textbf{} &
  \multicolumn{2}{c|}{\textbf{With AI}} &
  \multicolumn{3}{c|}{\textbf{Without AI}} \\ \hline
\textbf{LP} &
  \multicolumn{2}{l|}{\begin{tabular}[c]{@{}l@{}}\textbf{Food Quality:}\\ cheap good dessert bad breakfast decoration\\ overall bad food try another venue\\ good Thai food very happy meal\\ lazy service decent food won't return\\ \\ \textbf{Service and Cleanliness:}\\ self-service dirty restaurant won't visit again\\ good food nice people recommended visit\\ lazy service decent food won't return\end{tabular}} &
  \multicolumn{3}{l|}{\begin{tabular}[c]{@{}l@{}}\textbf{Food Quality:}\\ cheap good location dessert bad breakfast\\ good server ok pizza bad burger\\ nice people good food have games\\ tasty coconut soup and pad thai\\ good promotion friendly people tasty food\\ lazy service good food average pricing\\ \\ \textbf{Service and Cleanliness:}\\ poor service dirty live music avail\\ clean good service tasty reasonable price\\ lazy service good food average pricing\end{tabular}} \\ \hline
\textbf{SS} &
  \multicolumn{1}{l}{\begin{tabular}[c]{@{}l@{}}\textbf{Food Quality:}\\ cheap food\\ bad food\\ ok food\\ good food\\ expensive food\\ quality dropped\\ \\ \textbf{Service:}\\ good service\\ bad service\\ ok service\end{tabular}} &
  \begin{tabular}[c]{@{}l@{}}\textbf{Ambiance/Environment:}\\ bad decoration\\ good music\\ good entertainment\\ dirty\\ \\ \textbf{Others:}\\ good offer\\ \\ \end{tabular} &
  \multicolumn{1}{l}{\begin{tabular}[c]{@{}l@{}}\\\textbf{Food Quality:}\\ feels cheap\\ good food, service\\ neutral food\\ bad food\\ dislike burger set\\ cold, old fries\\ pricey pizza\\ okay pizza\\ good food people\\ coconut soup pad thai\\ same menu tried\\ liked coconut soup\\ coconut soup creamy\\ good pad thai\\ good peanuts, noodles\\ good chicken\\ good hot wings\\ lots of sauce\\ delicious sushi, affordable\\ affordable sushi\\ freshness and variety\\\end{tabular}} &
  \multicolumn{1}{l}{\begin{tabular}[c]{@{}l@{}}\textbf{Ambiance/Atmosphere:}\\ jaded decor\\ new orleans vibe\\ quiet, competent chef\\ \\ \\ \textbf{Service:}\\ poor service recovery (2)\\ decent server\\ lack of service\\ water not served\\ decent service\\ friendly staff\\ order mixup\\ \\ Recommendations and Reviews:\\ positive recommendation (2)\\ mixed review\end{tabular}} &
  \begin{tabular}[c]{@{}l@{}}\textbf{Customer Relationship:}\\ better previous experience\\ lost customer (mentioned twice)\\ purchase inconvenient\\ recent customer\\ overall satisfied\\ potentially lost customer\\ \\ \\ \textbf{Others:}\\ not clean\\ beer with brother\\ played nintendo\\ prefer fewer herbs\\ near hotel, convenient\\ promo good\\ returning for pizza\\ large party, hibachi\end{tabular} \\ \hline
\textbf{ME} &
  \multicolumn{1}{l}{\begin{tabular}[c]{@{}l@{}}\textbf{Food Quality:}\\ food tastes bad\\ food tastes normal\\ tasty\\ fresh food with huge variety\\ good place food and people\\ \\ \\ \textbf{Service:}\\ poor service\\ good service\\ cheap but poor food and service\end{tabular}} &
  \begin{tabular}[c]{@{}l@{}}\textbf{Pricing:}\\ expensive\\ worth\\ worth and tasty\\ expensive and tasty\\ tastes and feels cheap\\ \\ \\ \textbf{Others:}\\ will not try this again\\ unhygienic\\ quiet\end{tabular} &
  \begin{tabular}[c]{@{}l@{}}\textbf{Food Quality:}\\ cheap\\ good price location and dessert\\ bad burger\\ bad fries\\ pricey but ok pizza\\ good coconut soup\\ good pat thai\\ good hot wings\\ great sushi and reasonable price\\ fresh with huge variety\\ bad service good food but expensive\end{tabular} &
  \multicolumn{2}{l|}{\begin{tabular}[c]{@{}l@{}}\textbf{Service:}\\ bad services\\ poor service (mentioned twice)\\ good server\\ nice and friendly\\ clean environment and good service\\ great place, food and people\\ \\ \textbf{Attitude and others}\\ will not come again\\ dirty environment\\ good location\\ recommended\\ 1 for 1 offer\\ bad service good food but expensive\end{tabular}} \\ \hline
\end{tabular}}}
\end{table}

\subsection{RQ4: Impact on Perceived Trustworthiness and Helpfulness}
\label{sec:rq:perceived}

\subsubsection{\emph{\pt}} Results of users' self-reported trustworthiness are depicted in Table~\ref{tab:perceive-quality} and Figure~\ref{fig:helpfulness}. 

\paragraph{Code Granularity}
There is a significant main effect of \codeG on users' \pt to \cs ($F_{(2, 27)} = 3.449, p=.046$). The pairwise comparison revealed that the \pt to \cs under the \lo condition ($M=3.37$) surpassed that under the \mi condition ($M=2.67, p=.044$). No other pairwise differences were identified.

\paragraph{Text Granularity}
There is a significant main effect of \textG on \pt to \ra of the code suggestions ($F_{(2, 54)} = 3.512, p=.037$). Pairwise comparison showed that the \pt to \ra appeared to be superior in \sele ($M = 3.50$) compared to \sent ($M = 2.83, p< .01$). No other main effects were discerned.

\paragraph{Interaction effects}
No interaction effects were detected on \pt.

\begin{table}[!t]
\caption{Summary of Values of \pt and \he. All DVs are on a Likert scale from 1 to 5.}
\small
\label{tab:perceive-quality}
\scalebox{0.8}{\begin{tabular}{cccccc}
\hline
\multirow{2}{*}{\textbf{\begin{tabular}[c]{@{}c@{}}Factor1: Code\\ Granularity\end{tabular}}} & \multirow{2}{*}{\textbf{\begin{tabular}[c]{@{}c@{}}Factor2: Text\\ Granularity\end{tabular}}} & \multicolumn{3}{c}{\textbf{\emph{\pt}}} & \multirow{2}{*}{\textbf{\begin{tabular}[c]{@{}c@{}}Perceived Helpfulness\\ (M±S.D.)\end{tabular}}} \\ \cline{3-5}
 &  & \textbf{\begin{tabular}[c]{@{}c@{}}Confidence Score\\ (M±S.D.)\end{tabular}} & \textbf{\begin{tabular}[c]{@{}c@{}}Rank\\ (M±S.D.)\end{tabular}} & \textbf{\begin{tabular}[c]{@{}c@{}}Containing Ability\\ (M±S.D.)\end{tabular}} &  \\ \hline
\multirow{3}{*}{\begin{tabular}[c]{@{}c@{}}Short Codes\\ (1-3 words)\end{tabular}} & Sentence & 2.80 ± 0.92 & 3.10 ± 1.37 & 3.00 ± 1.25 & 3.60 ± 1.07 \\
 & Paragraph & 2.40 ± 0.97 & 3.10 ± 1.29 & 2.20 ± 1.34 & 2.30 ± 0.95 \\
 & Selective & 3.10 ± 0.88 & 3.80 ± 0.92 & 3.30 ± 1.16 & 3.70 ± 1.16 \\ \hline
\multirow{3}{*}{\begin{tabular}[c]{@{}c@{}}Long Codes\\ (4-6 words)\end{tabular}} & Sentence & 3.30 ± 0.94 & 3.10 ± 0.99 & 3.60 ± 0.97 & 3.80 ± 1.32 \\
 & Paragraph & 3.40 ± 0.84 & 3.60 ± 0.84 & 3.40 ± 1.17 & 4.30 ± 1.16 \\
 & Selective & 3.40 ± 1.17 & 3.60 ± 0.96 & 3.60 ± 0.84 & 3.80 ± 1.31 \\ \hline
\multirow{3}{*}{\begin{tabular}[c]{@{}c@{}}Mix Codes\\ (1-6 words)\end{tabular}} & Sentence & 2.40 ± 1.35 & 2.30 ± 0.95 & 2.40 ± 1.35 & 2.70 ± 1.16 \\
 & Paragraph & 3.10 ± 0.88 & 3.00 ± 1.05 & 3.40 ± 0.97 & 3.50 ± 0.71 \\
 & \textbf{Selective (baseline)} & \textbf{2.50 ± 0.97} & \textbf{3.10 ± 0.99} & \textbf{3.20 ± 1.14} & \textbf{1.40 ± 0.97} \\ \hline
\end{tabular}}
\end{table}

\subsubsection{Perceived Helpfulness}

\paragraph{Code Granularity} A significant main effect of \codeG on \ph were observed ($F_{(2, 27)} = 9.789, p<.001$).
The system's suggestions were deemed more helpful by users in the \lo condition ($M = 3.97$) than those in the \mi condition ($M = 2.53, p<.001$).
Likewise, in the \sh condition, the system was perceived as more helpful ($M = 3.20$) than in the \mi condition ($M = 2.53, p=.049$).

\paragraph{Text Granularity} No significant main effect on \ph was discerned.

\paragraph{Interaction effects}
A noteworthy interaction was detected between two factors on \ph of the system ($F_{(4, 54)} = 7.94, p<.001$). Particularly, under the \mi condition, the system was rated significantly more helpful in the \para condition compared to the \sele condition ($p < .001$).

\begin{figure}[!t]
    \centering
    \includegraphics[scale=0.5]{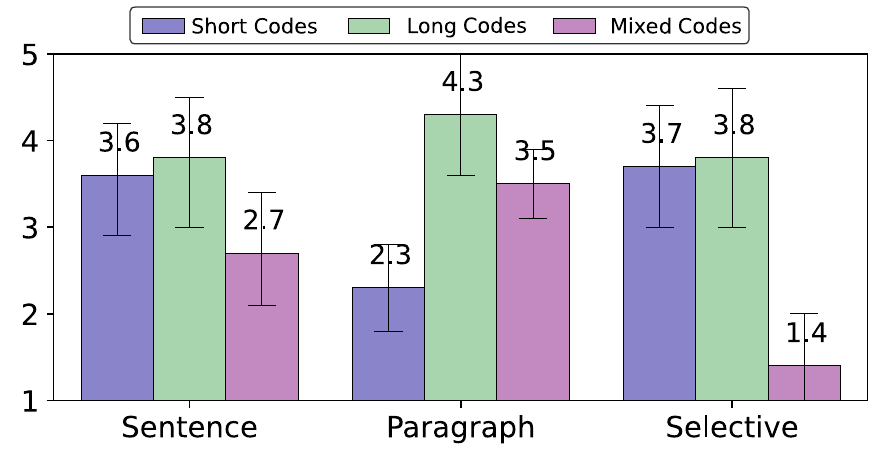}
    \caption{User's \he of code suggestions. Error bars show .95 confidence intervals. Y-axis represents 1-5 Likert score, where 1 represents a complete lack of helpfulness and 5 is the highest level of helpfulness.}
    \label{fig:helpfulness}
\end{figure}

\added{For descriptive statistics, the mean \he scores exceeds 3 (refer to Figure~\ref{fig:helpfulness}), albeit experiencing slight reductions under particular conditions such as \mi $\times$ \sele, \mi $\times$ \sent, and \sh $\times$ \para. We also observed that pairing \lo with a high level of \textG (\para) resulted in the highest mean \ph (4.3/5), significantly surpassing the scores in any of the other eight conditions. Unexpectedly, the baseline condition (\mi $\times$ \sele) results in the lowest \ph. Moreover, when \sh is paired with \para coding, it results in significantly diminished \ph, with users rating the system as not helpful. We delve deeper into this phenomenon from the perspective of task difficulty in Section \ref{sec:disc}.}


\subsection{RQ5: Impact on Subjective Preferences}
\label{sec:rq:subject}
In this section, we encapsulate the feedback conveyed by participants during and subsequent to the study. 

\subsubsection{User Preferred \emph{\sele}}
Greater control over the selection enabled participants to receive more accurate suggestions, which subsequently motivated them to choose suggestions more frequently. 
This is evidenced by the participant comment: \textit{"Because I can adjust the selection, then I think the way the numbers (confidence score) work...sometimes the one on the top is the one I want."} (P26, \mi and \sele).

Even though the length of text selections was similar between \sele and \sent, participants showed a preference for \sele, as articulated by P18: \textit{"The main difference is that for the sentence one, some sentences don't have meaning. But for selective, I could group the sentences with the same meaning together under one topic."} (\lo and \sele). 


\subsubsection{Imperfect AI Suggestions Still Contribute Value}
In many instances, participants found the suggested codes were close to their original ideas: \textit{"I find it relatively helpful. The recommended codes bear some similarity to what I had in mind, so I don't have to ponder excessively."} (P20, \lo and \para).

Whereas, even if the suggestions didn't always precisely align with the participants' requirements, they were still viewed as helpful. The suggested codes from the list inspired participants to combine existing codes to generate new ones and refine their own. The system's feature that allowed users to modify suggested codes was particularly appreciated:

\textit{"I think they [suggestions] are quite helpful. They kind of give you a hint about what you could write for the keywords or summaries."} (P28, \mi and \sent).

\subsubsection{Code Suggestions Promote Consistency}
Several participants highlighted that the suggestions aided them in maintaining consistency throughout the coding process:

\textit{"The recommendation list seems somewhat helpful because it enables me to apply a consistent metric when assessing these text streams. As a result, I can establish a bit more consistency between the texts as I formulate my codes."} (P22, \mi and \sent).

\subsubsection{\added{Too Long Text Selections (\emph{\para}) Presents Challenges}}
\added{Overall, participants indicated that AI would need to bolster its performance to meet their expectations. Specifically, a notable challenge inherent in AI is its struggle to capture nuanced information within the text:}

\textit{"The system failed to capture the context of sentences within the paragraph. At times, sentences were unrelated to one another - one might discuss good service while another addressed food, indicating different contexts within each review. Consequently, the system couldn't discern the nuances of individual sentences and provide accurate confidence scores."} (P39, \mi and \sent).

\section{Discussion}
\label{sec:disc}

\added{Our discussion first delves into an examination of task difficulty, outlining how various conditions were deemed more challenging than others. Following this, we unpack the diverse elements of trust between humans and \aiqc under evaluation. Lastly, we traverse through the significance of differing granularity conditions, especially in relation to their relevance in more realistic qualitative analysis scenarios.}

\subsection{Task Difficulty Across Conditions for \opencoding}
\label{sec:disc:difficulty}

\subsubsection{\added{Qualitative Open Coding: A Series of Distinct Tasks Rather than a Singular Whole}}

\added{In essence, our nine conditions simulate various levels of difficulty associated with \opencoding tasks. Our findings advance a nuanced understanding, positing that coding tasks can differ based on their intrinsic difficulty or the effort demanded. Some tasks may boost the model's performance, while others might hinder it. This viewpoint diverges from prior research in this domain, which might have predominantly treated \opencoding as a uniform, undifferentiated task, intending to devise a solitary method to facilitate it. This view has overlooked the inherent complexity of subjective tasks like qualitative coding, a scenario where human-AI interaction could play a pivotal role.}

\subsubsection{\added{Challenging \emph{\para} Conditions}}

\added{In \para, the units of text to be coded were longer compared to those in the \sent and \sele conditions. Moreover, based on subjective feedback, \para might have included various contradictory nuanced information.
Therefore, participants needed more time to decide on a code, leading us to infer that the coding task under the \para conditions was relatively more challenging. }

\added{However, according to the model performance data, an increase in context and text selection seems to enable the model to get higher precision. Conversely, a decrease in the text selection did not yield the same level of model performance.}

\added{At first glance, they may seem contradictory to each other. However, a paragraph may have a higher probability of matching the code, but it may also contain extraneous and even contradictory information not included in the ``highly matching code". Consequently, despite the possibility of paragraphs yielding higher model performance scores, they present a substantial challenge to both users and AI, particularly due to the nuanced information they may encapsulate.}

\subsubsection{The Complexity of \emph{\lo} Compared to \emph{\sh} and \emph{\mi}} 
The \lo conditions seemingly posed greater challenges for participants due to the requirement of a minimum number of words for each code, as indicated by the extended decision-making times. Conversely, participants found the \sh conditions less strenuous as they closely mirrored conventional coding scenarios, with code lengths similar to those in the \mi.

\subsection{\added{Trust Discrepancies Due to Varied Task Difficulties}}
\label{sec:disc:discrepancy}

\subsubsection{Higher Behavioral Trust for Simpler Tasks}
\label{sec:disc:behav}
In terms of \textG, our data reveal that users exhibit greater \bt, or reliance on AI, in simpler tasks at the \sele and \sent levels, as indicated by a 26\%+ suggestion selection rate. This contrasts with the harder \para tasks, which only saw an average selection rate of 16\%.

Significant individual differences also emerged. For instance, in \sele and \sent conditions, certain participants (P14 for \sele, P7 for \sent) ignored all suggestions, while others had high selection rates (73\% for P10 in \sele, 53\% for P17 in \sent). For the tougher \para conditions, an overwhelming 11 participants completely disregarded the system's suggestions, with the peak selection rate merely reaching 38\% (P25, P27, P28).


\subsubsection{Contrasting Behavioral Trust and Perceived Helpfulness in Complex Tasks}

Interestingly, while \bt was at its nadir in the \para setting, \ph was substantially high, and \pt was also considerable, especially for more complex tasks with longer AI suggestions. In particular, during tasks with longer codes (\lo $\times$ \para and \mi $\times$ \para), users, while finding individual suggestions inadequate for selection (thus the low \sr), still referenced them or made minor adjustments to construct their codes. This added flexibility enhanced their \ph, especially in the challenging task of \lo $\times$ \para, scoring 4.3/5 in \ph.


\subsection{\added{Over- and under-reliance on AIQCs}}
As discussed, different coding tasks resulted in varying reliance on the system. However, \aiqc should strike a balance between user exploration and AI assistance, without promoting either an excessive reliance or insufficient use of AI suggestions. Over-reliance can lead to shallow coding, while under-reliance may result in missed opportunities for valuable AI assistance. As such, striking the right balance is crucial for the effective use of AI in qualitative coding.

\subsubsection{Reasons for Under-reliance}
\added{The low \sr for \para tasks is primarily due to fewer coding units, resulting in fewer data points for model training. Typically, participants would only choose from the last few suggestions, with a total selection count of approximately 8, in comparison to around 35 in \sent and 20+ in \sele tasks. Thus, we anticipate that with more data points to train the model, the system reliance would increase.}

\added{For those scenarios where data points may not significantly increase, enabling users to edit their codes post-selection could offer indirect assistance (through the process of selection and then editing). Although these improvements might not be prominently reflected in \sr, they could elevate users' subjective experience, potentially bolster the system's trustworthiness, and encourage users to fully exploit the system \cite{banovic2023being}.}

\subsubsection{Over-reliance Risk}

\added{As detailed in section \ref{sec:rq:reliance_corr} and \ref{sec:rq:followup_study}, reliance could ostensibly reduce human effort by enabling users to re-utilize previous codes, causing a more focused coding. However, it also carries an over-reliance risk. Over-reliance might disrupt the delicate balance between focused coding and the generation of diverse coding outcomes, which could narrow the scope of interpretation and potentially reduce the depth and breadth of the coding process. Therefore, ensuring a balance between AI assistance and human input is key to maximizing both the efficiency and depth of qualitative coding. This balance should never be underestimated in the design of a truly trustworthy \aiqc, as opposed to a system that deceives users' trust without meriting it \cite{banovic2023being}.}

\subsection{Optimal Code Granularity Varies Between Users and AI}

Participants usually generate shorter codes when possible. The \mi, which allows them to create more specific and longer codes, resembles real-life open coding tasks more closely. The similar code lengths in \sh and \mi under both \sele and \sent conditions imply that participants aim to minimize their codes' length when given the option in the given study. Hence, \sh or \mi can be considered optimal code granularity for users.

Conversely, while users prefer to add shorter codes, they anticipate longer code suggestions from AI. This is evident in the consistently higher perceived helpfulness of \lo over other \textG conditions. We infer that longer AI suggestions enable more expressiveness, thereby reducing potential misinterpretations between users and AI.

Moreover, a discrepancy exists between human and AI preferences when it comes to text selection for coding. Participants generally favored labeling shorter selections that accommodated shorter codes, as indicated by the similar selection lengths in the \sh $\times$ \sele and \mi $\times$ \sele cases, where users selected only the critical single semantic elements for coding. On the other hand, AI favored a more comprehensive context for accurate code prediction, thus creating a divergence between human and AI inclinations. 
\added{In particular, the \sh $\times$ \para condition exemplified a considerable mismatch. Although it offers the optimal code length for users, the restriction of a three-word code for an extensive text led to a disconnect between the text and code, significantly increasing the task difficulty. This resulted in the lowest \sr of 11\% among users and lower \ph than neutral (2.3/5).}



In addition, users seemed to favor uniform AI suggestions, as indicated by their perception of \mi suggestions as less helpful than both \lo and \sh. The mix of long and short codes in the suggestion list under \mi might have led to information overload, making it challenging for users to locate useful information. Moreover, while the added flexibility, notably in \sele codes, could theoretically benefit the users, it seemed to inadvertently decrease the user's perceived helpfulness of the suggestions in \mi $\times$ \sele.


\subsection{Coding Strategies in Real Life}

\subsubsection{\emph{\sele} is Best for Coding}
Notably, we observed the highest levels of \bt in the \sele coding conditions. \sele coding most accurately emulates how a single researcher might begin to navigate data in a real-life scenario. They would select the most pertinent phrases and then generate a suitable label.  In practice, it is crucial to utilize various coding levels, alternating perspectives, and varying depths of understanding to produce a more comprehensive and diverse range of codes.

\subsubsection{\emph{\sent} for Collaborative Coding}
In fact, \sele and \sent coding scenarios share several similarities, notwithstanding certain notable differences. The variance between these two conditions is significantly less than that between them and the \para condition. While \sele may generally be the go-to granularity for coding, \sent level coding can be particularly beneficial for collaborative coding, where consistency between multiple users is required. This is especially important when computing inter-rater reliability scores, as it requires a straightforward, unambiguous text selection unit.

\subsubsection{\emph{\para} for Summarizing Long Texts}
\para may still prove valuable for users attempting to summarize lengthy texts in real coding scenarios. Opting for a \para approach assists in distilling entire pages into a few concise labels.

\section{Implications for Design}

\added{We propose several guidelines to cultivate appropriate reliance and foster a productive human-AI collaboration within the context of \aiqc.}

\subsection{\added{Fostering Trustworthiness during Under-reliance on \aiqc}}

\subsubsection{Offering Extensive and Modifiable Suggestions}

We observed that while users generally found less difficulty in creating \sh and \mi, \lo suggestions seem to be perceived as more beneficial and trustworthy. 
The utility of longer suggestions stems from their capacity to convey a wealth of information, thereby minimizing ambiguity and potentially delivering deeper meaning. This extensive nature also enables users to refine their code by editing suggestions, tailoring them to their unique requirements. This active participation makes users feel more in control, which could result in increased trust and system usage.

\subsubsection{Exploiting Larger Training Datasets}

We noted that some participants did not utilize AI suggestions during the \paraS tasks. This lack of use could be attributed to the reduced quantity and quality of the data used for training, resulting in initially subpar AI suggestions.

To address this concern, we propose the application of data augmentation techniques\footnote{\url{https://www.tensorflow.org/tutorials/images/data_augmentation}}, generating additional training data. Furthermore, if feasible, the integration of data from diverse users and sources could be beneficial for open coding. This recommendation aligns with the current trend towards a data-centric approach, as advocated in recent literature~\cite{eyuboglu2022dcbench, motamedi2021data, whang2021data}.

\subsubsection{Facilitating Open Coding Through Multifaceted Models}

We further recommend utilizing multiple models to generate code outputs from diverse perspectives. Rather than exclusively relying on text classification or topic modeling, we advocate for considering and integrating other methodologies, such as Generative AI like GPT \footnote{https://atlasti.com/ai-coding-powered-by-openai, https://openai.com/chatgpt}. By doing so, users can construct their codes under a wider umbrella of system assistance, thereby enabling more informed decision-making~\cite{feuston2021putting, jiang2021supporting}.
Moreover, the system could offer suggestions inspired by codes from other users, thus presenting an alternate view of the data.

\subsection{Mitigating Over-reliance to Prevent Shallow Codes}

We've recognized the potential for over-reliance in certain situations. \added{At times, AI that lacks sufficient trustworthiness could deceive users into considering it `trustworthy' \cite{banovic2023being}, leading to excessive reliance.} Below, we delve into several specific design strategies to mitigate this issue.


\subsubsection{Implementing a Delay in Suggestions Display upon Selection.}
The system could be designed to deliberately delay the display of suggestions or only present codes upon a user's request, ensuring they appear specifically when a user struggles to formulate a code \cite{buccinca2021trust}. This feature would afford the user sufficient time to contemplate an initial code, and subsequently ensure that the displayed suggestions align effectively with their requirements. 



\added{\subsubsection{Providing Explanations for AI Suggestions}
A promising strategy might be to present explanations alongside the code suggestions \cite{vasconcelos2023explanations}. For example, by displaying the original data from which the suggestions are derived, coders can compare and ascertain the appropriateness of coding the current data under a specific code. This approach not only encourages deeper thinking but also fosters appropriate reliance on the system.}
\section{Limitations and Future Work}
\label{limitation}

This work has limitations. 
\added{First, understanding the accuracy and overall performance of the model is crucial for gauging the system's effectiveness under varying conditions. Ideally, each text segment should have a corresponding ground truth value (such as 0 or 1) against which we can compare system recommendations to evaluate model performance.}

\added{To approximate this, we have considered each user's final code as a proximate `ground truth' for the specific text segment. This approximation is based on two assumptions: 1) the system merely plays an assisting role while the user remains the ultimate decision-maker; 2) all recommendations are derived from users' own coding history, enabling them to fully comprehend these suggestions. They can then decide whether to accept, modify, or reject these suggestions in order to make a final decision. Therefore, their assigned code for a text segment can thus be considered as a close `ground truth' representation for various users.} 

\added{Nevertheless, we understand that our approach only provides an estimation of the model's performance and may introduce \textbf{measurement errors}: 1) using the user's final code as `ground truth' presumes that the user's decisions are always accurate and ideal. However, users can make mistakes or demonstrate biases in their coding decisions. In such scenarios, the model's performance evaluation for a given text segment may be flawed, as the `ground truth' itself might not be correct; 2) interpreting the final user decision as `ground truth' could potentially inflate the model's performance metrics. Users often accept or make minor modifications to system suggestions, but this does not necessarily signify the model's recommendations were entirely accurate or the best available option. For example, in instances where users frequently adopt the model's recommendations despite them being merely "not bad" as opposed to the best, the performance assessment could be artificially elevated. As such, the model may appear to have a higher performance score, not necessarily because it offers the best suggestions, but because users tend to agree with its recommendations. This inflated performance metric could potentially misrepresent the model's true capacity to deliver optimal solutions across diverse contexts and user behaviors. }

\added{While we have accounted for this measurement error in our interpretation, future research could further investigate better ways to evaluate model performance, or establish a more suitable ground truth for subjective tasks such as qualitative coding.}


\added{Moreover, our choice to focus on codes of varying abstraction levels and specificity stems primarily from their representation of different user coding habits has been stated in Section \ref{sec:related_work:human_AI_interaction}.
In particular, to emulate these different levels of abstraction and interpretation, we opted for a simplistic albeit imperfect method for user operation. Codes of three words or less represent concise coding (short codes), those between four to six words signify verbose coding (long codes), and codes ranging from one to six words (mixed codes) represent natural coding. We selected these parameters for the study setup based on pilot tests conducted on our own materials prior to the formal study. 
However, we acknowledge that this classification has its shortcomings. For instance, the specific length of the codes is intimately linked to the domain of the coding material, and the delineation of codes across different levels of the factor remains unclear.
Looking forward, it would be promising to extend these results to various types of content, with the goal of gaining a more comprehensive understanding of the specific assistance and suggestions users truly need.
}

\added{Furthermore, there are also some limitations when assessing user trust (\pt) in \aiqc. For instance, users may struggle to differentiate their specific emotions and levels of trust towards individual components of the system (confidence score, rank, and containing ability), thereby influencing the overall evaluation of \pt. To address this, future research should invest more effort into developing more precise measures for trust evaluation. }

\added{Additionally, the selected parameters of limitation (1-3 words, 4-6 words, etc.) used for coding are, while simplistic, imperfect representations of user operations, mirroring the range from concise to lengthy and natural coding habits. The specific values, however, could vary significantly based on the coding material. Moreover, it's essential to motivate participants to execute tasks with greater efficiency, thereby achieving a more precise measure of decision-making time.
Future studies should further investigate these aspects, aiming to devise more general strategies for controlling and managing human-AI interaction habits.}

\added{\textbf{Overall, our primary objective in this work is to appeal to developers and researchers, underlining the importance of developing trustworthy \aiqc that fosters robust human-AI collaboration by taking into account the unique dynamics of human-AI interaction within qualitative coding.} It is critical to not only integrate advanced technologies into this domain but also to view \opencoding as a collection of different subtasks. Therefore, the design of various tools should aim to support the nuanced and varied coding tasks inherent within \opencoding. Furthermore, the potential risks for under-utilization (under-reliance) and over-reliance should be considered, as the former could result in the system being under-utilized, and the latter might lead to less insightful coding outcomes.}
\section{Conclusion}
\added{Issues concerning trust between humans and AI in \aiqc have been identified, but the exploration has remained limited. In this work, we explored how \emph{Code} and \emph{Text Granularity} could influence user trust and reliance in \aiqc by conducting a split-plot design study with 30 participants and a follow-up study with 6 participants. Our study highlighted that \opencoding, due to its unique human-AI interaction dynamics, should be approached as a composite of various subtasks. Each of these subtasks necessitates a tailored design. Our findings also indicate trust discrepancies stemming from varied subtask difficulties and illuminate the problems of over-reliance and under-reliance existing in different conditions. These results form a foundation for future research on the user trust, reliance, and utility of \aiqc. }

\bibliographystyle{ACM-Reference-Format}
\bibliography{paper/draft.bib}

\newpage

\appendix

\section{Results}

\subsection{\added{Model Performance}}
\label{appendix:result:model_performance}

\begin{table}[!htbp]
\caption{\added{A mixed two-way ANOVA result for Precision@5 of automatic evaluation of model performance, where $^{*}$ p < 0.05, $^{**}$ p < 0.01, $^{***}$ p < 0.001.}}
\scalebox{0.8}{\added{\begin{tabular}{|l|l|l|l|l|l|l|l|l|}
\hline
\textbf{Source} & \textbf{SS} & \textbf{DF1} & \textbf{DF2} & \textbf{MS} & \textbf{F} & \textbf{p-unc} & \textbf{np2} & \textbf{eps} \\ \hline
code\_granularity & 0.13 & 2 & 27 & 0.06 & 2.20 & 0.13 & 0.14 &  \\ \hline
text\_granularity & 0.15 & 2 & 54 & 0.08 & 5.80 & 0.01* & 0.18 & 0.90 \\ \hline
Interaction & 0.15 & 4 & 54 & 0.04 & 2.81 & 0.03* & 0.17 &  \\ \hline
\end{tabular}}}
\end{table}

\begin{table}[!htbp]
\caption{\added{Pairwise comparison results for Precision@5 of automatic evaluation of model performance, where $^{*}$ p < 0.05, $^{**}$ p < 0.01, $^{***}$ p < 0.001.}}
\scalebox{0.65}{\added{\begin{tabular}{|l|l|l|l|l|l|l|l|l|l|l|l|l|}
\hline
\textbf{code\_granularity} & \textbf{A} & \textbf{B} & \textbf{Paired} & \textbf{Parametric} & \textbf{T} & \textbf{dof} & \textbf{alternative} & \textbf{p-unc} & \textbf{p-corr} & \textbf{p-adjust} & \textbf{BF10} & \textbf{hedges} \\ \hline
- & Sentence & Paragraph & TRUE & TRUE & -1.02 & 29 & two-sided & 0.32 & 0.95 & bonf & 0.31 & -0.2 \\ \hline
- & Sentence & Selective & TRUE & TRUE & 2.54 & 29 & two-sided & 0.02* & 0.05* & bonf & 2.93 & 0.52 \\ \hline
- & Paragraph & Selective & TRUE & TRUE & 2.8 & 29 & two-sided & 0.01* & 0.03* & bonf & 4.91 & 0.69 \\ \hline
Long & Sentence & Paragraph & TRUE & TRUE & -0.54 & 9 & two-sided & 0.60 & 1.00 & bonf & 0.35 & -0.19 \\ \hline
Long & Sentence & Selective & TRUE & TRUE & 2.80 & 9 & two-sided & 0.02 & 0.19 & bonf & 3.49 & 0.93 \\ \hline
Long & Paragraph & Selective & TRUE & TRUE & 3.24 & 9 & two-sided & 0.01 & 0.09 & bonf & 6.15 & 1.15 \\ \hline
Mixed & Sentence & Paragraph & TRUE & TRUE & -1.24 & 9 & two-sided & 0.25 & 1.00 & bonf & 0.57 & -0.49 \\ \hline
Mixed & Sentence & Selective & TRUE & TRUE & 2.58 & 9 & two-sided & 0.03 & 0.27 & bonf & 2.63 & 1.30 \\ \hline
Mixed & Paragraph & Selective & TRUE & TRUE & 2.39 & 9 & two-sided & 0.04 & 0.37 & bonf & 2.05 & 1.07 \\ \hline
Short & Sentence & Paragraph & TRUE & TRUE & 0.10 & 9 & two-sided & 0.92 & 1.00 & bonf & 0.31 & 0.04 \\ \hline
Short & Sentence & Selective & TRUE & TRUE & -0.76 & 9 & two-sided & 0.47 & 1.00 & bonf & 0.39 & -0.20 \\ \hline
Short & Paragraph & Selective & TRUE & TRUE & -0.63 & 9 & two-sided & 0.55 & 1.00 & bonf & 0.36 & -0.24 \\ \hline
\end{tabular}}}
\end{table}

\begin{table}[!htbp]
\caption{\added{A mixed two-way ANOVA result for Recall@5 of automatic evaluation of model performance, where $^{*}$ p < 0.05, $^{**}$ p < 0.01, $^{***}$ p < 0.001.}}
\scalebox{0.8}{\added{\begin{tabular}{|l|l|l|l|l|l|l|l|l|}
\hline
\textbf{Source} & \textbf{SS} & \textbf{DF1} & \textbf{DF2} & \textbf{MS} & \textbf{F} & \textbf{p-unc} & \textbf{np2} & \textbf{eps} \\ \hline
code\_granularity & 0.27 & 2 & 27 & 0.13 & 3.08 & 0.06 & 0.19 &  \\ \hline
text\_granularity & 0.87 & 2 & 54 & 0.43 & 19.12 & $\leq0.001$*** & 0.41 & 0.86 \\ \hline
Interaction & 0.10 & 4 & 54 & 0.02 & 1.06 & 0.38 & 0.07 &  \\ \hline
\end{tabular}}}
\end{table}

\begin{table}[!htbp]
\caption{\added{Pairwise comparison results for Recall@5 of human evaluation of model performance, where $^{*}$ p < 0.05, $^{**}$ p < 0.01, $^{***}$ p < 0.001.}}
\scalebox{0.7}{\added{\begin{tabular}{|l|l|l|l|l|l|l|l|l|l|l|l|}
\hline
\textbf{A} & \textbf{B} & \textbf{Paired} & \textbf{Parametric} & \textbf{T} & \textbf{dof} & \textbf{alternative} & \textbf{p-unc} & \textbf{p-corr} & \textbf{p-adjust} & \textbf{BF10} & \textbf{hedges} \\ \hline
Sentence & Paragraph & TRUE & TRUE & -2.64 & 29 & two-sided & 0.01 & 0.04 & bonf & 3.57 & -0.64 \\ \hline
Sentence & Selective & TRUE & TRUE & 3.63 & 29 & two-sided & $\leq$0.001 & $\leq$0.001*** & bonf & 30.71 & 0.66 \\ \hline
Paragraph & Selective & TRUE & TRUE & 6.51 & 29 & two-sided & $\leq$0.001 & $\leq$0.001*** & bonf & 41140.00 & 1.40 \\ \hline
\end{tabular}}}
\end{table}

\begin{table}[!htbp]
\caption{\added{A mixed two-way ANOVA result for MAP@5 of automatic evaluation of model performance, where $^{*}$ p < 0.05, $^{**}$ p < 0.01, $^{***}$ p < 0.001.}}
\scalebox{0.76}{\added{\begin{tabular}{|l|l|l|l|l|l|l|l|l|l|l|l|l|}
\hline
\textbf{Source} & \textbf{SS} & \textbf{DF1} & \textbf{DF2} & \textbf{MS} & \textbf{F} & \textbf{p-unc} & \textbf{p-GG-corr} & \textbf{np2} & \textbf{eps} & \textbf{sphericity} & \textbf{W-spher} & \textbf{p-spher} \\ \hline
code\_granularity & 0.27 & 2 & 27 & 0.14 & 2.65 & 0.09 &  & 0.16 &  &  &  &  \\ \hline
text\_granularity & 1.12 & 2 & 54 & 0.56 & 20.30 & $\leq$0.001 & $\leq$0.001$^{***}$ & 0.43 & 0.72 & FALSE & 0.61 & 0.0009$^{***}$ \\ \hline
Interaction & 0.04 & 4 & 54 & 0.01 & 0.39 & 0.81 &  & 0.03 &  &  &  &  \\ \hline
\end{tabular}}}
\end{table}

\begin{table}[!htbp]
\caption{\added{Pairwise comparison results for MAP@5 of automatic evaluation of model performance, where $^{*}$ p < 0.05, $^{**}$ p < 0.01, $^{***}$ p < 0.001.}}
\scalebox{0.72}{\added{\begin{tabular}{|l|l|l|l|l|l|l|l|l|l|l|l|}
\hline
\textbf{A} & \textbf{B} & \textbf{Paired} & \textbf{Parametric} & \textbf{T} & \textbf{dof} & \textbf{alternative} & \textbf{p-unc} & \textbf{p-corr} & \textbf{p-adjust} & \textbf{BF10} & \textbf{hedges} \\ \hline
Sentence & Paragraph & TRUE & TRUE & -3.38 & 29 & two-sided & $\leq$0.01$^{**}$ & 0.01$^{**}$ & bonf & 17.22 & -0.81 \\ \hline
Sentence & Selective & TRUE & TRUE & 3.41 & 29 & two-sided & $\leq$0.01$^{**}$ & 0.01$^{**}$ & bonf & 18.61 & 0.58 \\ \hline
Paragraph & Selective & TRUE & TRUE & 6.28 & 29 & two-sided & $\leq$0.001$^{***}$ & $\leq$0.001$^{***}$ & bonf & 23300.00 & 1.34 \\ \hline
\end{tabular}}}
\end{table}

\begin{table}[!htbp]
\caption{\added{A mixed two-way ANOVA result for Precision@5 of human evaluation of model performance, where $^{*}$ p < 0.05, $^{**}$ p < 0.01, $^{***}$ p < 0.001.}}
\scalebox{0.75}{\added{\begin{tabular}{|l|l|l|l|l|l|l|l|l|}
\hline
\textbf{Source} & \textbf{SS} & \textbf{DF1} & \textbf{DF2} & \textbf{MS} & \textbf{F} & \textbf{p-unc} & \textbf{np2} & \textbf{eps} \\ \hline
code\_granularity & 0.07 & 2 & 27 & 0.04 & 1.46 & 0.25 & 0.10 &  \\ \hline
text\_granularity & 0.01 & 2 & 54 & 0.01 & 0.21 & 0.81 & 0.01 & 0.86 \\ \hline
Interaction & 0.06 & 4 & 54 & 0.02 & 0.51 & 0.73 & 0.04 &  \\ \hline
\end{tabular}}}
\end{table}

\begin{table}[!htbp]
\caption{\added{A mixed two-way ANOVA result for MAP@5 of human evaluation of model performance, where $^{*}$ p < 0.05, $^{**}$ p < 0.01, $^{***}$ p < 0.001.}}
\scalebox{0.75}{\added{\begin{tabular}{|l|l|l|l|l|l|l|l|l|l|l|l|l|}
\hline
\textbf{Source} & \textbf{SS} & \textbf{DF1} & \textbf{DF2} & \textbf{MS} & \textbf{F} & \textbf{p-unc} & \textbf{p-GG-corr} & \textbf{np2} & \textbf{eps} & \textbf{sphericity} & \textbf{W-spher} & \textbf{p-spher} \\ \hline
code\_granularity & 0.31 & 2 & 27 & 0.16 & 1.73 & 0.20 &  & 0.11 &  &  &  &  \\ \hline
text\_granularity & 1.44 & 2 & 54 & 0.72 & 37.00 & $\leq$0.001 & $\leq$0.001*** & 0.58 & 0.79 & FALSE & 0.73 & 0.01 \\ \hline
Interaction & 0.22 & 4 & 54 & 0.05 & 2.82 & 0.03 &  & 0.17 &  &  &  &  \\ \hline
\end{tabular}}}
\end{table}

\begin{table}[!htbp]
\caption{\added{Pairwise comparison results for MAP@5 of human evaluation of model performance, where $^{*}$ p < 0.05, $^{**}$ p < 0.01, $^{***}$ p < 0.001.}}

\scalebox{0.65}{\added{\begin{tabular}{|l|l|l|l|l|l|l|l|l|l|l|l|l|}
\hline
\textbf{code\_granularity} & \textbf{A} & \textbf{B} & \textbf{Paired} & \textbf{Parametric} & \textbf{T} & \textbf{dof} & \textbf{alternative} & \textbf{p-unc} & \textbf{p-corr} & \textbf{p-adjust} & \textbf{BF10} & \textbf{hedges} \\ \hline
- & Sentence & Paragraph & TRUE & TRUE & -7 & 29 & two-sided & $\leq0.001$ & $\leq0.001$*** & bonf & 300000 & -1 \\ \hline
- & Sentence & Selective & TRUE & TRUE & -4 & 29 & two-sided & $\leq0.001$ & 0.001*** & bonf & 80 & -0.6 \\ \hline
- & Paragraph & Selective & TRUE & TRUE & 5 & 29 & two-sided & $\leq0.001$ & $\leq0.001$*** & bonf & 300 & 0.8 \\ \hline
Long & Sentence & Paragraph & TRUE & TRUE & -7 & 9 & two-sided & $\leq0.001$ & $\leq0.001$*** & bonf & 500 & -2 \\ \hline
Long & Sentence & Selective & TRUE & TRUE & -3 & 9 & two-sided & 0.01 & 0.1 & bonf & 5 & -0.8 \\ \hline
Long & Paragraph & Selective & TRUE & TRUE & 3 & 9 & two-sided & 0.03 & 0.2 & bonf & 3 & 1 \\ \hline
Mixed & Sentence & Paragraph & TRUE & TRUE & -5 & 9 & two-sided & $\leq0.001$ & 0.006** & bonf & 60 & -2 \\ \hline
Mixed & Sentence & Selective & TRUE & TRUE & -2 & 9 & two-sided & 0.04 & 0.4 & bonf & 2 & -0.5 \\ \hline
Mixed & Paragraph & Selective & TRUE & TRUE & 4 & 9 & two-sided & 0.003 & 0.01** & bonf & 3 & 1 \\ \hline
Short & Sentence & Paragraph & TRUE & TRUE & -2 & 9 & two-sided & 0.06 & 0.5 & bonf & 2 & -0.6 \\ \hline
Short & Sentence & Selective & TRUE & TRUE & -2 & 9 & two-sided & 0.08 & 0.7 & bonf & 1 & -0.3 \\ \hline
Short & Paragraph & Selective & TRUE & TRUE & 1 & 9 & two-sided & 0.2 & 1 & bonf & 0.7 & 0.4 \\ \hline
\end{tabular}}}
\end{table}

\newpage

\subsection{\added{Coding Behavior, Decision Time}}
\label{sec:rq2_coding_behavior}


\begin{table}[!htbp]
\caption{\added{Pairwise comparison results for length of code, where $^{*}$ p < 0.05, $^{**}$ p < 0.01, $^{***}$ p < 0.001.}}
\scalebox{0.65}{\added{\begin{tabular}{|l|l|l|l|l|l|l|l|l|l|l|l|l|}
\hline
\textbf{text\_granularity} & \textbf{A} & \textbf{B} & \textbf{Paired} & \textbf{Parametric} & \textbf{T} & \textbf{dof} & \textbf{alternative} & \textbf{p-unc} & \textbf{p-corr} & \textbf{p-adjust} & \textbf{BF10} & \textbf{hedges} \\ \hline
- & Sentence & Paragraph & TRUE & TRUE & -5.64 & 29 & two-sided & \textless{}0.001 & \textless{}0.001*** & bonf & 4582.827 & -0.83 \\ \hline
- & Sentence & Selective & TRUE & TRUE & 0.47 & 29 & two-sided & 0.64 & 1 & bonf & 0.215 & 0.03 \\ \hline
- & Sentence & Selective & TRUE & TRUE & 0.47 & 29 & two-sided & 0.64 & 1 & bonf & 0.215 & 0.03 \\ \hline
- & Long & Mixed & FALSE & TRUE & 10.12 & 18 & two-sided & \textless{}0.001 & \textless{}0.001*** & bonf & 9.09E+05 & 4.33 \\ \hline
- & Mixed & Short & FALSE & TRUE & 8.22 & 18 & two-sided & \textless{}0.001 & \textless{}0.001*** & bonf & 5.38E+04 & 3.52 \\ \hline
Sentence & Long & Mixed & FALSE & TRUE & 12.61 & 18 & two-sided & \textless{}0.001 & \textless{}0.001*** & bonf & 2.29E+07 & 5.4 \\ \hline
Sentence & Mixed & Short & FALSE & TRUE & 3.35 & 18 & two-sided & 0 & 0.03* & bonf & 11.127 & 1.43 \\ \hline
Paragraph & Long & Mixed & FALSE & TRUE & 1.41 & 18 & two-sided & 0.18 & 1 & bonf & 0.788 & 0.6 \\ \hline
Paragraph & Mixed & Short & FALSE & TRUE & 12.07 & 18 & two-sided & \textless{}0.001 & \textless{}0.001*** & bonf & 1.17E+07 & 5.17 \\ \hline
Selective & Long & Mixed & FALSE & TRUE & 9.74 & 18 & two-sided & \textless{}0.001 & \textless{}0.001*** & bonf & 5.34E+05 & 4.17 \\ \hline
Selective & Mixed & Short & FALSE & TRUE & 2.48 & 18 & two-sided & 0.02 & 0.21 & bonf & 2.87 & 1.06 \\ \hline
\end{tabular}}}
\end{table}

\begin{table}[!htbp]
\caption{\added{Pairwise comparison results for length of selections, where $^{*}$ p < 0.05, $^{**}$ p < 0.01, $^{***}$ p < 0.001.}}
\scalebox{0.68}{\added{\begin{tabular}{|l|l|l|l|l|l|l|l|l|l|l|l|l|}
\hline
\textbf{text\_granularity} & \textbf{A} & \textbf{B} & \textbf{Paired} & \textbf{Parametric} & \textbf{T} & \textbf{dof} & \textbf{alternative} & \textbf{p-unc} & \textbf{p-corr} & \textbf{p-adjust} & \textbf{BF10} & \textbf{hedges} \\ \hline
Selective & Long & Mixed & FALSE & TRUE & 5.41 & 18 & two-sided & \textless{}0.001 & \textless{}0.001*** & bonf & 444.21 & 2.32 \\ \hline
Selective & Long & Short & FALSE & TRUE & 3.05 & 18 & two-sided & 0.007 & 0.062 & bonf & 6.81 & 1.31 \\ \hline
Selective & Mixed & Short & FALSE & TRUE & -0.94 & 18 & two-sided & 0.361 & 1.000 & bonf & 0.54 & -0.40 \\ \hline
\end{tabular}}}
\end{table}

\begin{table}[!htbp]
\caption{\added{Pairwise comparison results for number of selections, where $^{*}$ p < 0.05, $^{**}$ p < 0.01, $^{***}$ p < 0.001.}}
\scalebox{0.68}{\added{\begin{tabular}{|l|l|l|l|l|l|l|l|l|l|l|l|l|}
\hline
\textbf{text\_granularity} & \textbf{A} & \textbf{B} & \textbf{Paired} & \textbf{Parametric} & \textbf{T} & \textbf{dof} & \textbf{alternative} & \textbf{p-unc} & \textbf{p-corr} & \textbf{p-adjust} & \textbf{BF10} & \textbf{hedges} \\ \hline
Selective & Long & Mixed & FALSE & TRUE & -2.563 & 18 & two-sided & 0.020 & 0.117 & bonf & 3.214 & -1.098 \\ \hline
Selective & Long & Short & FALSE & TRUE & -2.603 & 18 & two-sided & 0.018 & 0.108 & bonf & 3.411 & -1.115 \\ \hline
Selective & Mixed & Short & FALSE & TRUE & -0.223 & 18 & two-sided & 0.826 & 1.000 & bonf & 0.404 & -0.095 \\ \hline
\end{tabular}}}
\end{table}

\begin{table}[!htbp]
\caption{\added{A mixed two-way ANOVA results for Decision Time, where $^{*}$ p < 0.05, $^{**}$ p < 0.01, $^{***}$ p < 0.001.}}
\scalebox{0.8}{\added{\begin{tabular}{|l|l|l|l|l|l|l|l|l|}
\hline
\textbf{Source} & \textbf{SS} & \textbf{DF1} & \textbf{DF2} & \textbf{MS} & \textbf{F} & \textbf{p-unc} & \textbf{np2} & \textbf{eps} \\ \hline
code\_granularity & 6675.63 & 2 & 24 & 3337.82 & 11.13 & <0.001*** & 0.48 &  \\ \hline
text\_granularity & 4951.46 & 2 & 48 & 2475.73 & 10.13 & <0.001*** & 0.30 & 0.83 \\ \hline
Interaction & 126.06 & 4 & 48 & 31.52 & 0.13 & 0.971 & 0.01 &  \\ \hline
\end{tabular}}}
\end{table}

\begin{table}[!htbp]
\caption{\added{Pairwise comparison results for Decision Time, where $^{*}$ p < 0.05, $^{**}$ p < 0.01, $^{***}$ p < 0.001.}}
\scalebox{0.7}{\added{\begin{tabular}{|l|l|l|l|l|l|l|l|l|l|l|l|}
\hline
\textbf{A} & \textbf{B} & \textbf{Paired} & \textbf{Parametric} & \textbf{T} & \textbf{dof} & \textbf{alternative} & \textbf{p-unc} & \textbf{p-corr} & \textbf{p-adjust} & \textbf{BF10} & \textbf{hedges} \\ \hline
Long & Mixed & FALSE & TRUE & 4.297 & 16 & two-sided & 0.001 & 0.002** & bonf & 47.53 & 1.93 \\ \hline
Long & Short & FALSE & TRUE & 3.276 & 16 & two-sided & 0.005 & 0.014* & bonf & 8.97 & 1.47 \\ \hline
Mixed & Short & FALSE & TRUE & -0.555 & 16 & two-sided & 0.586 & 1.000 & bonf & 0.46 & -0.25 \\ \hline
Paragraph & Selective & TRUE & TRUE & 2.883 & 26 & two-sided & 0.008 & 0.023* & bonf & 5.75 & 0.66 \\ \hline
Paragraph & Sentence & TRUE & TRUE & 4.208 & 26 & two-sided & 0.000 & 0.001** & bonf & 108.38 & 0.95 \\ \hline
Selective & Sentence & TRUE & TRUE & 1.635 & 26 & two-sided & 0.114 & 0.342 & bonf & 0.66 & 0.33 \\ \hline
\end{tabular}}}
\end{table}

\newpage

\subsection{\added{Selecting Rate}}

\begin{table}[!htbp]
\caption{\added{A mixed two-way ANOVA results for Selecting Rate, where $^{*}$ p < 0.05, $^{**}$ p < 0.01, $^{***}$ p < 0.001.}}
\scalebox{0.8}{\added{\begin{tabular}{|l|l|l|l|l|l|l|l|l|}
\hline
\textbf{Source} & \textbf{SS} & \textbf{DF1} & \textbf{DF2} & \textbf{MS} & \textbf{F} & \textbf{p-unc} & \textbf{np2} & \textbf{eps} \\ \hline
code\_granularity & 0.042 & 2 & 27 & 0.021 & 0.391 & 0.680 & 0.028 &  \\ \hline
text\_granularity & 0.403 & 2 & 54 & 0.202 & 15.838 & <0.001*** & 0.370 & 0.881 \\ \hline
Interaction & 0.141 & 4 & 54 & 0.035 & 2.766 & 0.036 & 0.170 &  \\ \hline
\end{tabular}}}
\end{table}

\begin{table}[!htbp]
\caption{\added{Pairwise comparison results for Selecting Rate, where $^{*}$ p < 0.05, $^{**}$ p < 0.01, $^{***}$ p < 0.001.}}
\scalebox{0.6}{\added{\begin{tabular}{|l|l|l|l|l|l|l|l|l|l|l|l|l|}
\hline
\textbf{code\_granularity} & \textbf{A} & \textbf{B} & \textbf{Paired} & \textbf{Parametric} & \textbf{T} & \textbf{dof} & \textbf{alternative} & \textbf{p-unc} & \textbf{p-corr} & \textbf{p-adjust} & \textbf{BF10} & \textbf{hedges} \\ \hline
\textbf{-} & Paragraph & Selective & TRUE & TRUE & -4.515 & 29 & two-sided & 0.001 & <0.001*** & bonf & 266.048 & -0.921 \\ \hline
- & Paragraph & Sentence & TRUE & TRUE & -3.507 & 29 & two-sided & 0.001 & 0.004** & bonf & 23.215 & -0.708 \\ \hline
- & Selective & Sentence & TRUE & TRUE & 2.186 & 29 & two-sided & 0.037 & 0.111 & bonf & 1.521 & 0.337 \\ \hline
Long & Paragraph & Selective & TRUE & TRUE & -1.205 & 9 & two-sided & 0.259 & 1 & bonf & 0.552 & -0.327 \\ \hline
Long & Paragraph & Sentence & TRUE & TRUE & -0.816 & 9 & two-sided & 0.436 & 1 & bonf & 0.408 & -0.197 \\ \hline
Long & Selective & Sentence & TRUE & TRUE & 0.832 & 9 & two-sided & 0.427 & 1 & bonf & 0.412 & 0.137 \\ \hline
Mixed & Paragraph & Selective & TRUE & TRUE & -2.439 & 9 & two-sided & 0.037 & 0.337 & bonf & 2.197 & -0.804 \\ \hline
Mixed & Paragraph & Sentence & TRUE & TRUE & -1.417 & 9 & two-sided & 0.190 & 1 & bonf & 0.674 & -0.544 \\ \hline
Mixed & Selective & Sentence & TRUE & TRUE & 1.140 & 9 & two-sided & 0.284 & 1 & bonf & 0.521 & 0.444 \\ \hline
Short & Paragraph & Selective & TRUE & TRUE & -4.685 & 9 & two-sided & 0.001 & 0.010** & bonf & 36.507 & -1.572 \\ \hline
Short & Paragraph & Sentence & TRUE & TRUE & -4.653 & 9 & two-sided & 0.001 & 0.011* & bonf & 35.185 & -1.531 \\ \hline
Short & Selective & Sentence & TRUE & TRUE & 1.629 & 9 & two-sided & 0.138 & 1 & bonf & 0.84 & 0.464 \\ \hline
\end{tabular}}}
\end{table}

\subsection{\added{Perceived Helpfulness}}

\begin{table}[!htbp]
\caption{\added{A mixed two-way ANOVA results for Perceived Helpfulness, where $^{*}$ p < 0.05, $^{**}$ p < 0.01, $^{***}$ p < 0.001.}}
\scalebox{0.8}{\added{\begin{tabular}{|l|l|l|l|l|l|l|l|l|}
\hline
\textbf{Source} & \textbf{SS} & \textbf{DF1} & \textbf{DF2} & \textbf{MS} & \textbf{F} & \textbf{p-unc} & \textbf{np2} & \textbf{eps} \\ \hline
code\_granularity & 30.87 & 2 & 27 & 15.43 & 9.79 & 0.001** & 0.420 &  \\ \hline
text\_granularity & 3.20 & 2 & 54 & 1.60 & 1.53 & 0.225 & 0.054 & 0.941 \\ \hline
Interaction & 33.13 & 4 & 54 & 8.28 & 7.94 & <0.001*** & 0.370 &  \\ \hline
\end{tabular}}}
\end{table}

\begin{table}[!htbp]
\caption{\added{Pairwise comparison results for Perceived Helpfulness, where $^{*}$ p < 0.05, $^{**}$ p < 0.01, $^{***}$ p < 0.001.}}
\scalebox{0.58}{\added{\begin{tabular}{|l|l|l|l|l|l|l|l|l|l|l|l|l|l|}
\hline
\textbf{Contrast} & \textbf{code\_granularity} & \textbf{A} & \textbf{B} & \textbf{Paired} & \textbf{Parametric} & \textbf{T} & \textbf{dof} & \textbf{alternative} & \textbf{p-unc} & \textbf{p-corr} & \textbf{p-adjust} & \textbf{BF10} & \textbf{hedges} \\ \hline
code\_granularity & - & Long & Mixed & FALSE & TRUE & 4.42 & 18 & two-sided & 0.000 & 0.001** & bonf & 72.836 & 1.892 \\ \hline
code\_granularity & - & Long & Short & FALSE & TRUE & 2.17 & 18 & two-sided & 0.044 & 0.131 & bonf & 1.855 & 0.929 \\ \hline
code\_granularity & - & Mixed & Short & FALSE & TRUE & -2.29 & 18 & two-sided & 0.034 & 0.103 & bonf & 2.177 & -0.980 \\ \hline
\begin{tabular}[c]{@{}l@{}}code\_granularity * \\ text\_granularity\end{tabular} & Long & Paragraph & Selectively & TRUE & TRUE & 1.34 & 9 & two-sided & 0.213 & 1.000 & bonf & 0.626 & 0.386 \\ \hline
\begin{tabular}[c]{@{}l@{}}code\_granularity * \\ text\_granularity\end{tabular} & Long & Paragraph & Sentence & TRUE & TRUE & 0.79 & 9 & two-sided & 0.453 & 1.000 & bonf & 0.399 & 0.386 \\ \hline
\begin{tabular}[c]{@{}l@{}}code\_granularity * \\ text\_granularity\end{tabular} & Long & Selectively & Sentence & TRUE & TRUE & 0.00 & 9 & two-sided & 1.000 & 1.000 & bonf & 0.309 & 0.000 \\ \hline
\begin{tabular}[c]{@{}l@{}}code\_granularity * \\ text\_granularity\end{tabular} & Mixed & Paragraph & Selectively & TRUE & TRUE & 6.68 & 9 & two-sided & 0.000 & 0.001** & bonf & 310.034 & 2.376 \\ \hline
\begin{tabular}[c]{@{}l@{}}code\_granularity * \\ text\_granularity\end{tabular} & Mixed & Paragraph & Sentence & TRUE & TRUE & 1.71 & 9 & two-sided & 0.121 & 1.000 & bonf & 0.923 & 0.798 \\ \hline
\begin{tabular}[c]{@{}l@{}}code\_granularity * \\ text\_granularity\end{tabular} & Mixed & Selectively & Sentence & TRUE & TRUE & -2.75 & 9 & two-sided & 0.022 & 0.202 & bonf & 3.267 & -1.167 \\ \hline
\begin{tabular}[c]{@{}l@{}}code\_granularity * \\ text\_granularity\end{tabular} & Short & Paragraph & Selectively & TRUE & TRUE & -2.94 & 9 & two-sided & 0.016 & 0.148 & bonf & 4.167 & -1.266 \\ \hline
\begin{tabular}[c]{@{}l@{}}code\_granularity * \\ text\_granularity\end{tabular} & Short & Paragraph & Sentence & TRUE & TRUE & -3.07 & 9 & two-sided & 0.013 & 0.119 & bonf & 4.944 & -1.228 \\ \hline
\begin{tabular}[c]{@{}l@{}}code\_granularity * \\ text\_granularity\end{tabular} & Short & Selectively & Sentence & TRUE & TRUE & 0.26 & 9 & two-sided & 0.798 & 1.000 & bonf & 0.318 & 0.086 \\ \hline
\end{tabular}}}
\end{table}

\end{document}